\documentclass[11pt]{article}
\pdfoutput=1
\usepackage{amsmath,amssymb,amsfonts,dsfont}
\usepackage[citebordercolor={.8 .8 1},urlbordercolor ={.8 .8 1},pdfstartview=FitV]{hyperref}

\numberwithin{equation}{section}

\begin{document}
\begin{center}

\vspace{1cm} { \Large {\bf Thermodynamics of Higher Spin Black Holes in AdS$_3$}}

\vspace{1.1cm}
Jan de Boer and Juan I. Jottar

\vspace{0.7cm}

{\it Institute for Theoretical Physics, University of Amsterdam,\\
Science Park 904, Postbus 94485, 1090 GL Amsterdam, The Netherlands}

{\tt J.deBoer@uva.nl, J.I.Jottar@uva.nl} \\

\vspace{1.5cm}

\end{center}

\begin{abstract}
\noindent We discuss the thermodynamics of recently constructed three-dimensional higher spin black holes in $SL(N,\mathds{R})\times SL(N,\mathds{R})$ Chern-Simons theory with generalized asymptotically-anti-de Sitter boundary conditions. From a holographic perspective, these bulk theories are dual to two-dimensional CFTs with $\mathcal{W}_{N}$ symmetry algebras, and the black hole solutions are dual to thermal states with higher spin chemical potentials and charges turned on. Because the notion of horizon area is not gauge-invariant in the higher spin theory, the traditional approaches to the computation of black hole entropy must be reconsidered. One possibility, explored in the recent literature, involves demanding the existence of a partition function in the CFT, and consistency with the first law of thermodynamics. This approach is not free from ambiguities, however, and in particular different definitions of energy result in different expressions for the entropy.  In the present work we show that there are natural definitions of the thermodynamically conjugate variables that follow from careful examination of the variational principle, and moreover agree with those obtained via canonical methods. Building on this intuition, we derive general expressions for the higher spin black hole entropy and free energy which are written entirely in terms of the Chern-Simons connections, and are valid for both static and rotating solutions. We compare our results to other proposals in the literature, and provide a new and efficient way to determine the generalization of the Cardy formula to a situation with higher spin charges.  

%\vspace{30pt}
%\centerline{Draft version of \today}
\end{abstract}

\pagebreak

\setcounter{page}{1}
\setcounter{equation}{0}
\tableofcontents

\pagebreak

%%%%%%%%%%%%%%%%%%%%%%%%%%%%%%%%%%%%%%%%%%%%%%%%%%%%%%%%%%%%%%%%%%%%%%%%%%%%%%%%
\section{Introduction}
%%%%%%%%%%%%%%%%%%%%%%%%%%%%%%%%%%%%%%%%%%%%%%%%%%%%%%%%%%%%%%%%%%%%%%%%%%%%%%%%
The recent surge in the study of higher spin theories and their holographic duals has raised several interesting puzzles. Partly, the interest in this class of theories originates in the desire of exploring the anti-de Sitter(AdS)/Conformal Field Theory(CFT) correspondence of \cite{Maldacena:1997re,Gubser:1998bc,Witten:1998qj} in a regime of parameters where the dual field theory is not necessarily strongly-coupled, and consequently the bulk gravitational theory does not reduce to classical (super-)gravity. Understanding and testing the holographic correspondence in different regimes is a theoretically appealing prospect, and departing from classical gravity presents us with the equally interesting challenge of extending the holographic dictionary to encompass these setups.

Perhaps the most widely studied example of higher spin holographic duality is rooted in the conjecture \cite{Klebanov:2002ja} of Klebanov and Polyakov relating critical $O(N)$ vector models in the large-$N$ limit and the higher spin Fradkin-Vasiliev theory in AdS$_{4}$ \cite{Fradkin:1987ks,Fradkin:1986qy}. Unfortunately, the formulation of interacting non-linear higher spin theories in four dimensions requires considerable technical machinery, and one may then hope to count with simplified setups where similar questions can be posed and addressed. One such possibility entails considering higher spin theories in AdS$_{3}$. Indeed, in three dimensions it is possible to build a consistent non-linear theory of gravity interacting with a finite number of higher spin fields \cite{Blencowe:1988gj}, as opposed to their higher-dimensional cousins where the infinite tower of higher spin fields is kept. Furthermore, the higher spin AdS$_{3}$ theories with spins $s \leq N$ can be cast in the form of Chern-Simons gauge theory with gauge group $SL(N,\mathds{R})\times SL(N,\mathds{R})\,$. Much in the same way that standard Einstein gravity with AdS$_{3}$ boundary conditions gives rise to an asymptotic symmetry group consisting of two copies of the Virasoro algebra \cite{Brown:1986nw}, the analysis of asymptotic symmetries in the Chern-Simons formulation of the higher spin theory \cite{Henneaux:2010xg,Campoleoni:2010zq,Gaberdiel:2010ar} reveals that they correspond to two-dimensional CFTs with (classical) $\mathcal{W}_{N}$ symmetry algebras \cite{Zamolodchikov:1985wn}.\footnote{This connection is not very surprising given the relation between Chern-Simons theories and current algebras on the one hand, and the relation between $\mathcal{W}_{N}$ algebras and the Hamiltonian reduction of current algebras \cite{Bershadsky:1989mf} on the other, see e.g. the discussion in section~3.2 of \cite{deBoer:1998ip}.}

A question of considerable interest from the AdS/CFT perspective (and in itself), is the construction of black hole solutions dual to thermodynamic equilibrium CFT states with higher spin charges and chemical potentials turned on. This problem was first addressed in \cite{Gutperle:2011kf,Ammon:2011nk,Castro:2011fm,Gaberdiel:2012yb} (see \cite{Ammon:2012wc} for a recent review). In particular it was found that the notion of horizon area is not gauge-invariant in the higher spin theory, and one must then reconsider the traditional methods of evaluating the black hole entropy such as the Bekenstein-Hawking formula. In \cite{Gutperle:2011kf} it was proposed to evaluate the entropy indirectly by demanding the existence of a well-defined partition function and consistency with the first law of thermodynamics. Other (in principle compatible) approaches involve the computation of the free energy and entropy directly from the Euclidean Chern-Simons action \cite{Banados:2012ue}, the extension of Wald's entropy formula \cite{Campoleoni:2012hp}, and the use of canonical methods \cite{Perez:2012cf,Perez:2013xi}. The indirect approaches for evaluating the entropy are however not free from ambiguities; for example, different definitions of energy result in different expressions for the entropy.

In the present work we argue that a careful examination of the variational principle in Lorentzian and Euclidean signature provides a natural way of defining the thermodynamically conjugate variables, namely the sources and the expectation values (EVs) of operators in the dual theory, and consequently the energy, entropy, and other quantities of interest. Our approach is similar in spirit to that of \cite{Banados:2012ue}, and extends it to include rotating solutions. We fill some gaps in the existing literature by working in the formalism where the chemical potential conjugate to the stress tensor is included as the modular parameter of the boundary torus in the Euclidean formulation of the theory, and provide general expressions written entirely in terms of the Chern-Simons connections. The latter feature is appropriate to the topological character of the bulk theory, and implies that our general expressions do not rely on specific details of the solutions (other than universal gauge choices). Moreover, our formulas are valid for any embedding of the gravitational $sl(2,\mathds{R})$ factor into $sl(N,\mathds{R})$, and apply to both static and rotating solutions. 

We should perhaps emphasize the general strategy at this point. To completely define Chern-Simons theory, we need not only the Chern-Simons action, but also a choice of boundary conditions plus a choice of boundary terms. If the variation of Chern-Simons action with suitable boundary terms is of the form $\sum_i O_i \,\delta J_i\,$, then one naturally interprets the $J_i$ as sources and the $O_i$ as expectation values. Moreover, the boundary conditions should be such that the values of $J_i$ at the boundary are fixed while the $O_i$ are allowed to fluctuate. The Euclidean action, evaluated with the same boundary terms, can then be viewed as a free energy which is a function of the sources $J_i\,$. By performing a Legendre transformation we obtain the entropy as a function of $O_i\,$. In principle, there can be different choices of boundary terms and boundary conditions that give rise to physically inequivalent theories. There can also be choices which are related to each other through a field redefinition. Ultimately, we would like to match any such description in Chern-Simons theory to a dual CFT and identify the precise relation between $J_i\,$, $O_i\,$, and the CFT variables.
This can be quite subtle, even more so in the presence of sources for irrelevant operators, and we will return to this point in the discussion section. 

This paper is organized as follows. In section \ref{CS gravity} we provide a brief introduction to the Chern-Simons formulation of standard three-dimensional gravity in the presence of a negative cosmological constant, and display the BTZ black hole solution in this language. In section \ref{HS gravity} we review some of the recent work in the $N>2$ case and discuss the structure of black hole solutions in Lorentzian signature, in the formalism where the source for the stress tensor is included explicitly in the Chern-Simons connections. Given the structure of the solutions, we introduce appropriate boundary terms that supplement the Chern-Simons action in order to achieve a well-defined Dirichlet variational principle. In section \ref{section: Thermo} we continue the black hole solutions and variational principle to Euclidean signature, and carefully discuss the transition to the formalism employed in most of the recent literature, where the temperature and angular momentum are incorporated via the periodicity of the boundary torus. We then provide explicit formulas for the energy and entropy of black hole solutions, which are written entirely in terms of the connection components and are therefore widely applicable. We conclude in section \ref{sec: conclusions} with an application of our formalism to a few explicit examples including the principal embedding spin-3 black hole constructed in \cite{Gutperle:2011kf,Ammon:2011nk}, and a comparison to the results obtained with other approaches. We will also discuss the connection to computations done in conformal field theory, and gather various other loose ends.

%%%%%%%%%%%%%%%%%%%%%%%%%%%%%%%%%%%%%%%%%%%%%%%%%%%%%%%%%%%%%%%%%%%%%%%%%%%%%%%%
\section{Brief review of $\text{AdS}_{3}$ gravity as a Chern-Simons theory}\label{CS gravity}
%%%%%%%%%%%%%%%%%%%%%%%%%%%%%%%%%%%%%%%%%%%%%%%%%%%%%%%%%%%%%%%%%%%%%%%%%%%%%%%%
 In $(2+1)$ dimensions, Einstein gravity with negative cosmological constant can be written as a Chern-Simons theory with gauge group $G \simeq SL(2,\mathds{R})\times SL(2,\mathds{R})\,$. This topological formulation was first introduced in \cite{Achucarro:1987vz}, and its quantum aspects subsequently discussed in \cite{Witten:1988hc} (see \cite{Witten:2007kt} for a modern review). More precisely, in three dimensions one can combine the dreibein $e^{a}$ and the dual spin connection
\begin{equation}\label{definition dual spin connection}
\omega^{a} \equiv \frac{1}{2}\epsilon^{abc}\omega_{bc}\quad \Leftrightarrow \quad \omega_{ab} = -\epsilon_{abc}\,\omega^{c}\, ,
\end{equation}
\noindent into $SL(2,\mathds{R})$ gauge connections $A$, $\bar{A}$ defined as
\begin{equation}
A =  \omega +\frac{e}{\ell}\quad \mbox{and}\quad \bar{A} =  \omega - \frac{e}{\ell}\,.
\end{equation}
\noindent Here, $\ell$ is a length scale set by the cosmological constant (i.e. the AdS$_{3}$ radius) and $\omega \equiv \omega^{a}J_{a}\,$, $e \equiv e^{a}J_{a}\,$, where the generators $J_{a}$ obey the $sl(2,\mathds{R}) \simeq so(2,1)$ algebra $\left[J_{a},J_{b}\right] = \epsilon_{abc}\,\eta^{cd}J_{d} = \epsilon_{ab}^{\hphantom{ab}c}J_{c}\,
$.

Defining the Chern-Simons form $CS(A)$ as
\begin{equation}
CS(A) = A\wedge dA + \frac{2}{3}A\wedge A\wedge A\,,
\end{equation}

\noindent  one observes that the combination $\mbox{Tr}\Bigl[CS(A) - CS(\bar{A})\Bigr]$ reproduces the Einstein-Hilbert Lagrangian, up to a boundary term. The precise relation is 
\begin{align}\label{CS action}
I_{CS} \equiv{}&
 \frac{k}{4\pi y_{R} }\int_{M} \mbox{Tr}\Bigl[CS(A) - CS(\bar{A})\Bigr]
 \\
={}&
 \frac{1}{16\pi G_{3}}\left[\int_{M} d^{3}x\sqrt{|g|}\left(\mathcal{R} + \frac{2}{\ell^{2}}\right) -\int_{\partial M}\omega^{a}\wedge e_{a}\right],
 \nonumber
\end{align}

\noindent  where $G_{3}$ is the three-dimensional Newton constant, $y_{R}$ is a representation-dependent normalization constant defined through $\mbox{Tr}\left[J_{a}J_{b}\right] = (y_{R}/2)\eta_{ab}\,$ (with conventions such that $y_{R}=1$ in the fundamental representation of $so(2,1)$), and 
%[\textbf{Note: look ahead at \eqref{kcs} and \eqref{central charge} for definitions in the general case}]
\begin{equation}
k = \frac{\ell}{4G_{3}}\,.
\end{equation}

\noindent In particular, Einstein's equations amount to the flatness of the connections: $F = dA + A\wedge A =0\,$, $\bar{F} = d\bar{A} + \bar{A}\wedge \bar{A} = 0\,$. As first shown by Brown and Henneaux \cite{Brown:1986nw}, boundary conditions in the three-dimensional gravity theory can be chosen such that the asymptotic symmetry group corresponds to two copies of the Virasoro algebra with central charge $c =  6k = 3\ell/(2G_{3})$ (see \cite{Coussaert:1995zp} for a derivation of this fact in the Chern-Simons formulation). These boundary conditions define what is commonly referred to as AdS$_{3}$ asymptotics, and provide a concrete realization of the AdS/CFT correspondence. 

  Let us consider the Chern-Simons theory on a (Lorentzian) three-dimensional manifold $M$ with topology $\mathds{R}\times D\,$, where the $\mathds{R}$ factor is associated with the time direction and $D$ is a two dimensional manifold with boundary $\partial D \simeq S^{1}$. It is customary to introduce coordinates $(\rho,t,\varphi)$ on $M$, where $\rho$ is a radial coordinate and the constant-radius surfaces (in particular the boundary $\partial M$ at $\rho \to \infty$) have the topology of a cylinder.  We will denote the $sl(2,\mathds{R})$ generators by $\{\Lambda^{0},\Lambda^{\pm}\}$, with 
\begin{equation}
\left[\Lambda^{\pm},\Lambda^{0}\right]=\pm \Lambda^{\pm}\,,\quad \left[\Lambda^{+},\Lambda^{-}\right]=2\Lambda^{0}\,.
\end{equation}  

\noindent Using the gauge freedom of the Chern-Simons theory, one can parameterize the space of asymptotically AdS$_{3}$ (AAdS$_{3}$) solutions with a \textit{flat} boundary metric on $\partial M$ by \cite{Banados:1998gg}
\begin{align}\label{general form connections}
A
=
 b^{-1} db + b^{-1}a\,b\,,\qquad 
\bar{A} 
=
 b\,db^{-1} +b\,\bar{a}\,b^{-1}\, ,
\end{align}
\noindent with $b = b(\rho) = e^{\rho \Lambda^{0}}$ and 
\begin{equation}
%\begin{aligned}\label{sl2r connections}
%a
%&=
% \left(L_{1} - \left(\frac{2\pi}{k}\right)\mathcal{L}(x^{+})L_{-1}\right)dx^{+}\,,
%\\
% \bar{a}
% &=
%  \left(-L_{-1}+ \left(\frac{2\pi}{k}\right)\bar{\mathcal{L}}(x^{-})L_{1}\right)dx^{-}\,,
%  \end{aligned}
\begin{aligned}\label{sl2r connections}
a
&=
 \left(\Lambda^{+} -\left(\frac{2\pi}{k}\right)\mathcal{L}(x^{+})\Lambda^{-}\right)dx^{+}\,,
\\
 \bar{a}
 &=
  -\left(\Lambda^{-}- \left(\frac{2\pi}{k}\right)\bar{\mathcal{L}}(x^{-})\Lambda^{+}\right)dx^{-}\,,
  \end{aligned}
\end{equation}

\noindent  where the light-cone coordinates $x^{\pm}$ are given by $x^{\pm} = t/\ell \pm \varphi\,$. In particular, the global AdS$_{3}$ solution corresponds to constant $\mathcal{L}$, $\bar{\mathcal{L}}$ given by $\mathcal{L}_{AdS_{3}} = \bar{\mathcal{L}}_{AdS_{3}}=-\frac{k}{8\pi}\,$, and the BTZ black hole \cite{Banados:1992wn} with mass $M$ and angular momentum $J$ is obtained with
\begin{equation}
\begin{aligned}\label{BTZ L and Lbar}
\mathcal{L}_{BTZ} &= \frac{1}{4\pi}\left(M\ell  - J\right) =  \frac{k}{2}\frac{\pi\ell^{2}}{\beta_{-}^{2}}\,
\\
\bar{\mathcal{L}}_{BTZ} &= \frac{1}{4\pi}\left(M\ell  + J\right) = \frac{k}{2}\frac{\pi\ell^{2}}{\beta_{+}^{2}}\, ,
%\label{BTZ L and Lbar 2}
%\end{align}
\end{aligned}
\end{equation}
\noindent  where $\beta_{\pm} = 1/T_{\pm}\,$ are the inverse chiral temperatures. From the point of view of the dual CFT$_{2}\,$, $\mathcal{L}$ and $\bar{\mathcal{L}}$ are seen to correspond to the stress tensor zero modes (see \cite{Kraus:2006wn} for an excellent review of the AdS$_{3}/$CFT$_{2}$ correspondence).

%%%%%%%%%%%%%%%%%%%%%%%%%%%%%%%%%%%%%%%%%%%%%%%%%%%%%%%%%%%%%%%%%%%%%%%%%%%%%%%%
\section{Higer spin theories in $\text{AdS}_{3}$}\label{HS gravity}
%%%%%%%%%%%%%%%%%%%%%%%%%%%%%%%%%%%%%%%%%%%%%%%%%%%%%%%%%%%%%%%%%%%%%%%%%%%%%%%%
Having interpreted AdS$_{3}$ Einstein gravity as an $SL(2,\mathds{R})\times SL(2,\mathds{R})$ Chern-Simons theory, it is natural to generalize the construction by promoting the gauge group to $SL(N,\mathds{R})\times SL(N,\mathds{R})\,$. When $N\geq 3\,$, this theory describes the non-linear interactions of gravity coupled to a finite tower of fields of spin $s \leq N$ \cite{Blencowe:1988gj}. The asymptotic symmetry analysis for these cases was performed in \cite{Henneaux:2010xg,Campoleoni:2010zq} (see also \cite{deBoer:1991jc,DeBoer:1992vm} for early work). Requiring that the asymptotic symmetries still include the Virasoro algebras already present in the pure gravity case ($N=2$), it was found that suitable boundary conditions for the higher spin connections are
\begin{gather}\label{AAdS3 bcs}
A_{-}=0\,,\qquad A_{\rho} = b^{-1}(\rho)\partial_{\rho}b(\rho)\,,\qquad
 A - A_{AdS_{3}} \xrightarrow[\rho \to \infty]{\phantom{\rho \to \infty}} \mathcal{O}(1)\,,
\end{gather}

\noindent where $A_{AdS_{3}}$ corresponds to the global $AdS_{3}$ solution and  $ b(\rho) = e^{\rho \Lambda^{0}}$ as before. With this choice the asymptotic symmetries are given by the so-called $\mathcal{W}_{N}$ algebras \cite{Zamolodchikov:1985wn}, which correspond to non-linear extensions of the Virasoro algebra. The form of the above boundary conditions uses ideas from the so-called Drinfeld-Sokolov reduction \cite{Drinfeld:1984qv}: roughly speaking, starting with an affine current algebra based on a simple Lie group one imposes constraints that reduce the phase space, resulting in the corresponding symmetry algebra (see \cite{Campoleoni:2011hg} for details). From this perspective, the Brown-Henneaux boundary conditions that define AdS$_{3}$ asymptotics in standard three-dimensional gravity correspond to the reduction of a $sl(2,\mathds{R})\times sl(2,\mathds{R})$ current algebra that leaves behind left- and right-moving Virasoro algebras. For general $N$, the boundary condition $(A - A_{AdS_{3}}) \xrightarrow[\rho \to \infty]{\phantom{\rho \to \infty}} \mathcal{O}(1)$ implements first order constraints (which generate gauge transformations) with respect to the current algebra: these allow to retain the Virasoro symmetries and enhance them to the non-linear $\mathcal{W}_{N}$ algebra on the reduced phase space \cite{Campoleoni:2010zq,Henneaux:2010xg}. 

The precise matter content of the gravitational theory, and hence the spectrum and symmetries of the dual CFT, depends on how the $SL(2,\mathds{R})$ subgroup associated to the gravity sector is embedded into $SL(N,\mathds{R})\,$ (see e.g. \cite{Bais:1990bs,deBoer:1993iz}, and \cite{Castro:2012bc} for a review); different embeddings are characterized by how the fundamental representation of $sl(N,\mathds{R})$ decomposes into $sl(2,\mathds{R})$ representations. These branching rules are classified by the different (integer) partitions of $N\,$ \cite{Dynkin:1957um}. For example, let us consider the fundamental representation $\mathbf{3}_{3}$ of $sl(3,\mathds{R})$. Denoting the $(2j +1)$-dimensional representation of $sl(2,\mathds{R})$ by $\mathbf{(2j+1)}_{2}\,$, we have three equivalence classes of embeddings characterized by the decompositions $\mathbf{3}_{3} \simeq \mathbf{3}_{2}\,$,  $\mathbf{3}_{3} \simeq \mathbf{2}_{2}\oplus \mathbf{1}_{2}\,$, $\mathbf{3}_{3} \simeq 3\cdot \mathbf{1}_{2}\,$ corresponding to partitions $\{0+3\}$, $\{1+2\}$ and $\{1+1+1\}\,$, respectively. The embedding characterized by $\mathbf{3}_{3} \simeq \mathbf{3}_{2}$ is dubbed ``principal embedding"; in general, principal embeddings are characterized by the fact that the fundamental representation becomes an irreducible representation of the embedded algebra. The branching rule  $\mathbf{3}_{3} \simeq 3\cdot \mathbf{1}_{2}\,$ is an example of ``trivial embedding", where only singlets appear in the decomposition of the fundamental representation. The second non-trivial embedding obtained from  $\mathbf{3}_{3} \simeq \mathbf{2}_{2}\oplus \mathbf{1}_{2}\,$ is called ``diagonal embedding", because the embedded $sl(2,\mathds{R})$ has a block-diagonal form in $sl(3,\mathds{R})\,$. 

Given the branching of the fundamental representation one can deduce the branching of the $\left(N^{2}-1\right)$-dimensional adjoint representation (see e.g. \cite{Bais:1990bs,deBoer:1992sy}). This determines the decomposition of the algebra itself and hence the spectrum. For example, in the principal embedding one finds $\text{adj}_{N} = \mathbf{3}_{2} \oplus \mathbf{5}_{2} \oplus \ldots \oplus \mathbf{(2N-1)}_{2}\,$. In other words, in the principal embedding the $sl(N,\mathds{R})$ algebra decomposes into $N-1$ representations with spins ranging from $1$ to $N-1\,$. From the point of view of the bulk theory, these correspond to bulk fields of spin 2 (the metric) and a tower of higher spin fields with spins $3,\ldots,N\,$. A general feature is that the conformal weight of the operators furnishing the representation in the boundary theory is obtained by adding one to the $sl(2,\mathds{R})$ spin (see e.g. \cite{Ammon:2011nk}). Then, in addition to the stress tensor (quasi-primary of weight two),  in the principal embedding one finds conformal primaries of weight $3,4\ldots, N\,$. Similarly, in the diagonal embedding one has $\text{adj}_{N} = \mathbf{3}_{2} \oplus 2(N-2)\cdot \mathbf{2}_{2} \oplus (N-2)^{2} \cdot \mathbf{1}_{2}\,$. Therefore, besides the stress tensor, the spectrum in the diagonal embedding includes currents of weight $1$ and $3/2\,$. Whenever charges fields are present, there is always a consistent truncation where they are taken to be equal to zero. When applied to the diagonal embedding, the spin $3/2$ fields are put equal to zero while the weight one currents are truncated to the diagonal subset: from the bulk point of view, the theory in the diagonal embedding contains a truncation to AdS$_{3}$ gravity coupled to $U(1)^{2(N-2)}$ gauge fields.  As discussed in \cite{Castro:2011fm}, this feature makes it particularly simple to construct black hole solutions in the diagonal embedding, inasmuch as they reduce to BTZ black holes charged under Abelian holonomies. 

In particular, in the $SL(3,\mathds{R})\times SL(3,\mathds{R})$ case the bulk theory in the principal embedding contains gravity non-linearly coupled to a spin-3 field. As shown in \cite{Henneaux:2010xg,Campoleoni:2010zq}, the corresponding asymptotic symmetry algebra then consists of two copies of the Zamolodchikov $\mathcal{W}_{3}$ algebra \cite{Zamolodchikov:1985wn}, with the central charge taking the same value as in the standard Brown-Henneaux calculation for asymptotically AdS$_{3}$ spacetimes \cite{Brown:1986nw}, i.e. $c = 6k = 3\ell/(2G_{3})$. According to the general discussion above, in addition to the stress tensor this algebra includes primary operators of dimensions $(3,0)$ and  $(0,3)$. We have also learned that for $N=3$ there is only one other non-trivial inequivalent embedding, namely the diagonal embedding. The solutions in this case asymptote to an AdS$_{3}$ vacuum with different radius, and the corresponding asymptotic symmetry algebra is identified with the so-called $\mathcal{W}_{3}^{(2)}$ or Polyakov-Bershadsky algebra \cite{Henneaux:2010xg,Campoleoni:2010zq,Ammon:2012wc}. In agreement with the general discussion above, apart from the stress tensor, this algebra contains weight-$3/2$ primary operators, as well as a weight one current. The central charge in the $\mathcal{W}_{3}^{(2)}$ case is given by $\hat{c} = c/4 = 3k/2\,$. 
 
We will write the coefficient of the Chern-Simons action in general as $k_{cs}/(4\pi)\,$; matching with the normalization of the Einstein-Hilbert action then requires
\begin{equation}\label{kcs}
k_{cs} = \frac{k}{2\mbox{Tr}\left[\Lambda^{0}\Lambda^{0}\right]}
\end{equation}

\noindent where $k$ is the level of the $SL(2)$ Chern-Simons theory that is contained in the $SL(N)$ Chern-Simons theory through the choice of $SL(2,\mathds{R})$ embedding, and so that, in terms of the level $k_{cs}$ of the $SL(N)$ theory, the central charge in the dual CFT is given by
\begin{equation}\label{central charge}
c  = 12 k_{cs}\text{Tr}\left[\Lambda^{0}\Lambda^{0}\right].
\end{equation}

\noindent Traces in the above equations are traces in the fundamental representation. We also notice that for fixed $k_{cs}$ the central charge changes for different embeddings. By the same token, in the general case the metric is constructed from the dreibein as  \cite{Castro:2011iw,Ammon:2012wc}
\begin{equation}
g_{\mu\nu} = \frac{1}{\text{Tr}\left[\Lambda^{0}\Lambda^{0}\right]}\text{Tr}\left[e_{\mu}e_{\nu}\right].
\end{equation}

%%%%%%%%%%%%%%%%%%%%%%%%%%%%%%%%%%%%%%%%%%%%%%%
\subsection{Turning on sources: black hole solutions}\label{Lorentzian bhs}
%%%%%%%%%%%%%%%%%%%%%%%%%%%%%%%%%%%%%%%%%%%%%%%
Since black holes are dual to states in thermodynamic equilibrium, black holes solutions carrying higher spin charges must also include chemical potentials which are the thermodynamic conjugate of the higher spin charges. In the holographic context, these chemical potentials source higher spin operators that deform the dual CFT. We will begin our discussion of black hole solutions in Lorentzian signature and carefully continue to the Euclidean formulation in the next section. In Lorentzian signature we choose to include the chemical potential conjugate to the stress tensor explicitly in the connections. In the Euclidean case one can instead include the source for the stress tensor as the modular parameter of the boundary torus, and we will discuss the technical differences in both approaches. 

The radial dependence of the black hole connections can be gauge-fixed to be of the form \eqref{general form connections}, so we focus on the $\rho$-independent connections $a$, $\bar{a}\,$. The general structure of the black hole solutions consistent with our discussion of boundary conditions is then
\begin{align}\label{DS connections}
a ={}&
 \Bigl(\Lambda^{+} + Q\Bigr)dx^{+} + \Bigl(M + \ldots \Bigr)dx^{-}
 \\
\bar{a} ={}&
 \Bigl(-\Lambda^{-} + \bar{Q}\Bigr)dx^{-} + \Bigl(\bar{M} + \ldots \Bigr)dx^{+}\,.
\label{DS connections 2}
\end{align}

\noindent We adopt the convention that the highest (lowest) weights in $a_{+}$ ($\bar{a}_{-}$) are linear in the charges (i.e. the EVs), and the lowest (highest) weights in $a_{-}$ ($\bar{a}_{+}$) are linear in the chemical potentials (i.e. the sources). In other words, the components of the matrices $Q$ and $\bar{Q}$ are linear in the charges and satisfy 
\begin{equation}
\left[\Lambda^{-},Q\right] = \left[\Lambda^{+},\bar{Q}\right] = 0\,,
\end{equation}

\noindent while the matrices $M$, $\bar{M}$ are linear in the chemical potentials and satisfy
\begin{equation}
\left[\Lambda^{+},M\right] = \left[\Lambda^{-},\bar{M}\right] = 0\,.
\end{equation}

\noindent The dots in \eqref{DS connections}-\eqref{DS connections 2} represent terms that are fixed by the equations of motion (e.g. $\left[a_{+},a_{-}\right]=\left[\bar{a}_{+},\bar{a}_{-}\right]=0$ for constant connections), and are in general non-linear in the charges. As part of our definition of charges and chemical potentials, in the principal embedding we demand
\begin{align}\label{mu Q terms}
\text{Tr}\Bigl[\left(a_{+} - \Lambda^{+}\right)a_{-}\Bigr] ={}& \text{Tr}\bigl[Q\,a_{-}\bigr] =  \sum_{j =2}^{N}\mu_{j}Q_{j}
\\
\text{Tr}\Bigl[\left(\bar{a}_{-} + \Lambda^{-}\right)\bar{a}_{+}\Bigr] ={}& \text{Tr}\bigl[\bar{Q}\,\bar{a}_{+}\bigr] = \sum_{j =2}^{N}\bar{\mu}_{j}\bar{Q}_{j}\,,
\label{mu Q terms 2}
\end{align}

\noindent where $Q_{j}$ is the charge associated with the operator of conformal weight $j\,$. In the principal embedding, the matrix $Q$ ($\bar{Q}$) introduced above can be easily written down as a sum over the highest (lowest) weight generator in each multiplet of spin $s=j-1=1\,,2\ldots\,N-1$ (see e.g. \cite{Castro:2011iw,Castro:2012bc}). As indicated above, the chemical potential $\mu_{2}$ conjugate to the stress tensor is included explicitly as a lowest weight in the connection. 

The extension of \eqref{mu Q terms}-\eqref{mu Q terms 2} to other embeddings is straightforward: for example, in the truncation of the diagonal embedding to gravity plus Abelian gauge fields we would instead require
\begin{align}\label{mu Q terms diag}
 \text{Tr}\left[Q\,a_{-}\right] ={}&
  \mu_{2}Q_{2} + \sum_{i =1}^{N-2}\mu^{(i)}_{1}Q^{(i)}_{1}
\\
 \text{Tr}\left[\bar{Q}\,\bar{a}_{+}\right] ={}&
  \bar{\mu}_{2}\bar{Q}_{2}+\sum_{i =1}^{N-2}\bar{\mu}^{(i)}_{1}\bar{Q}^{(i)}_{1}\,,
\label{mu Q terms 2 diag}
\end{align}

\noindent where $Q^{(i)}_{1}$, $\bar{Q}^{(i)}_{1}$ are the $U(1)$ charges associated with the weight-one currents. Besides the usual freedom of simultaneously rescaling $\mu_{j} \to \lambda \mu_{j}$ and $Q_{j} \to Q_{j}/\lambda\,$, the conditions \eqref{mu Q terms}-\eqref{mu Q terms 2} (or \eqref{mu Q terms diag}-\eqref{mu Q terms 2 diag}) still leave some ambiguity in the definition of the chemical potentials. We will return to this issue in section \ref{subsec:entropy}, where we discuss the entropy of black hole solutions. 

As a concrete example, let us write down the \textit{constant} Lorentzian solution for the $N=3$ theory in the principal embedding. Using the same conventions for the $sl(3,\mathds{R})$ generators as in \cite{Gutperle:2011kf}, we easily find that the constant connection that obeys our definitions is given by
\begin{equation}
a =  \left(\begin{array}{ccc}
0 & \frac{Q_{2}}{2} & Q_{3} \\ 
1 & 0 & \frac{Q_{2}}{2} \\ 
0 & 1 & 0
\end{array} \right)dx^{+} + \left(\begin{array}{ccc}
-\frac{\mu_{3}Q_{2}}{6} & \frac{\mu_{2}Q_{2}}{2} +\mu_{3}Q_{3} & \mu_{2}Q_{3} + \frac{\mu_{3}Q_{2}^{2}}{4} \\ 
\mu_{2} & \frac{\mu_{3}Q_{2}}{3} & \frac{\mu_{2}Q_{2}}{2} +\mu_{3}Q_{3} \\ 
\mu_{3} & \mu_{2} & -\frac{\mu_{3}Q_{2}}{6}
\end{array} \right)dx^{-}
\end{equation}
%\begin{equation}
%a_{+} = \left(\begin{array}{ccc}
%0 & \frac{Q_{2}}{2} & Q_{3} \\ 
%1 & 0 & \frac{Q_{2}}{2} \\ 
%0 & 1 & 0
%\end{array} \right)
%\end{equation}
%
%\noindent and 
%
%\begin{equation}
%a_{-}=\left(\begin{array}{ccc}
%-\frac{\mu_{3}Q_{2}}{6} & \frac{\mu_{2}Q_{2}}{2} +\mu_{3}Q_{3} & \mu_{2}Q_{3} + \frac{\mu_{3}Q_{2}^{2}}{4} \\ 
%\mu_{2} & \frac{\mu_{3}Q_{2}}{3} & \frac{\mu_{2}Q_{2}}{2} +\mu_{3}Q_{3} \\ 
%\mu_{3} & \mu_{2} & -\frac{\mu_{3}Q_{2}}{6}
%\end{array} \right)\,,
%\end{equation}

\noindent with a similar expression for the $\bar{a}$ connection.  The fact that some of the components of $a_{+}$ are fixed to be equal to $1$ is a manifestation of the constraints that reduce the phase space leaving behind the $\mathcal{W}_{N}$ algebras, which include the Virasoro symmetries. Consequently, we refer to \eqref{DS connections}-\eqref{DS connections 2} as the Drinfeld-Sokolov form of the connections. As we have emphasized, in Lorentzian signature the range of the coordinates is fixed and it is natural to include the chemical potential $\mu_{2}$ as a lowest weight in $a_{-}\,$. Different choices are possible in the Euclidean setup, and we will specify how this affects the identification of chemical potentials and charges.

It is worth mentioning that proper black hole solutions must satisfy additional smoothness conditions that can be phrased in terms of the holonomy of the connections around the contractible cycle in the Euclidean formulation \cite{Gutperle:2011kf,Ammon:2011nk}. These conditions relate the charges and chemical potentials in a way consistent with thermodynamic integrability, a point that will be crucial for the thermodynamic considerations in section \ref{section: Thermo}. 

%%%%%%%%%%%%%%%%%%%%%%%%%%%%%%%%%%%%%%%%%%%%%%%
\subsection{Lorentzian variational principle}
%%%%%%%%%%%%%%%%%%%%%%%%%%%%%%%%%%%%%%%%%%%%%%%
As in our discussion of the BTZ solution in section \ref{CS gravity}, we consider the Lorentzian Chern-Simons theory on a three-dimensional manifold $M$ with topology $\mathds{R}\times D\,$, where the $\mathds{R}$ factor is associated with the time direction and $\partial D \simeq S^{1}\,$. When defined on a manifold with boundary, the action 
\begin{align}\label{CS action}
I_{CS} =
 \frac{k_{cs}}{4\pi}\int_{M} \mbox{Tr}\Bigl[CS(A) - CS(\bar{A})\Bigr]
\end{align}

\noindent must be supplemented with suitable boundary terms in order to obtain a well defined variational principle. The total action will then be of the form
\begin{equation}\label{full action}
I = I_{CS} + I_{Bdy}\,,
\end{equation}

\noindent where $I_{Bdy}$ is the required boundary term. Given that the radial dependence of the solutions enters through a gauge transformation as in \eqref{general form connections}, the explicit factors of $b(\rho)$ largely drop out and we can phrase the discussion in terms of the $\rho$-independent connections $a$, $\bar{a}\,$. We first notice that the variation of $I_{CS}$, evaluated on-shell, yields 
\begin{align}\label{CS action variation}
\left.\delta I_{CS}\right|_{os}  
={}&
- \frac{k_{cs}}{4\pi}\int_{\partial M}\text{Tr}\Bigl[a\wedge \delta a - \bar{a}\wedge \delta \bar{a}\Bigr]
\\
 ={}&
 - \frac{k_{cs}}{2\pi}\int_{\partial M}d^{2}x\,\text{Tr}\Bigl[a_{+} \delta a_{-} - a_{-}\delta a_{+} -\bar{a}_{+} \delta \bar{a}_{-} + \bar{a}_{-}\delta \bar{a}_{+} \Bigr],
\end{align}

\noindent where we chose the orientation as $d^{2}x \equiv (1/2)dx^{-}\wedge dx^{+} = dt\,d\varphi\,$. We propose that the natural boundary term that is appropriate to the Drinfeld-Sokolov form of the connection is\footnote{In terms of the full connections $A$, $\bar{A}$, the appropriate boundary term is of the form  $\int d^{2}x\,\text{Tr}\left[\left(A_{+} -2b^{-1}(\rho) \Lambda^{+}b(\rho)\right)A_{-}\right]$ and similarly for the barred connection.}
%\begin{align}
%I_{bdy} ={}& -\frac{k}{4\pi y_{R}} \int_{\partial M} d^{2}x\, \text{Tr}\left[\left(a_{+} -2 \Lambda^{+}\right)a_{-}\right] 
%\nonumber\\
%&-\frac{k}{4\pi y_{R}}\int_{\partial M} d^{2}x\,  \text{Tr}\left[\left(\bar{a}_{-} +2 \Lambda^{-}\right)\bar{a}_{+} \right].
%\end{align}
\begin{align}\label{Lorentzian boundary term}
I_{Bdy} = -\frac{k_{cs}}{2\pi} \int_{\partial M} d^{2}x\, \text{Tr}\Bigl[\left(a_{+} -2 \Lambda^{+}\right)a_{-}\Bigr] 
-\frac{k_{cs}}{2\pi}\int_{\partial M} d^{2}x\,  \text{Tr}\Bigl[\left(\bar{a}_{-} +2 \Lambda^{-}\right)\bar{a}_{+} \Bigr].
\end{align}
\noindent The on-shell variation of the full action $I$ is then of the form
\begin{align}
\left.\delta\Bigl( I_{CS}+I_{Bdy}\Bigr)\right|_{os} = - \frac{k_{cs}}{2\pi}\int_{\partial M}d^{2}x\,\text{Tr}\Bigl[2\left(a_{+}-\Lambda^{+}\right)\delta a_{-} + 2\left(\bar{a}_{-} + \Lambda^{-}\right)\delta \bar{a}_{+}\Bigr].
\end{align}

\noindent Plugging in the solutions \eqref{DS connections}-\eqref{DS connections 2} in Drinfeld-Sokolov form and using \eqref{mu Q terms}-\eqref{mu Q terms 2}, in the principal embedding we obtain
\begin{align}
\left.\delta\Bigl( I_{CS}+I_{Bdy}\Bigr)\right|_{os} ={}&
 - \frac{k_{cs}}{\pi}\int_{\partial M}d^{2}x\,\text{Tr}\Bigl[Q\delta a_{-} + \bar{Q}\delta \bar{a}_{+}\Bigr]
 \\
  ={}&
  - \frac{k_{cs}}{\pi}\int_{\partial M}d^{2}x\,\sum_{j=2}^{N}\Bigl(Q_{j}\delta \mu_{j} + \bar{Q}_{j}\delta \bar{\mu}_{j}\Bigr).
\end{align}

\noindent Taking into account the different spectrum (c.f. \eqref{mu Q terms diag}-\eqref{mu Q terms 2 diag} for example), similar expressions are obtained in non-principal embeddings. We conclude that the action principle \eqref{full action} with the boundary term given by \eqref{Lorentzian boundary term} is indeed appropriate to the Dirichlet problem (fixed sources). It is worth emphasizing that this result holds even for solutions consisting of non-constant connections, because equations \eqref{DS connections}-\eqref{mu Q terms 2} defining the lowest/highest weight structure of the solutions continue to hold in this case.

%%%%%%%%%%%%%%%%%%%%%%%%%%%%%%%%%%%%%%%%%%%%%%%%%%%%%%%%%%%%%%%%%%%%%%%%%%%%%%%%
\section{Thermodynamics of higher spin black holes}\label{section: Thermo}
%%%%%%%%%%%%%%%%%%%%%%%%%%%%%%%%%%%%%%%%%%%%%%%%%%%%%%%%%%%%%%%%%%%%%%%%%%%%%%%%
In order to discuss the thermodynamics of the higher spin black holes, we will now study the structure of the solutions and the variational principle in Euclidean signature, where the bulk manifold has the topology of a solid torus. As we have emphasized, one could work with coordinates of fixed periodicity and explicitly introduce the chemical potential conjugate to the stress tensor as a lowest weight in the connection, as we did when discussing the Lorentzian solutions. In fact, in the fixed-periodicity formalism the partition function of the higher spin theory with arbitrary sources on the complex plane was computed long ago in \cite{deBoer:1991jc,DeBoer:1992vm}, where the answer was given in terms of Wess-Zumino-Witten (WZW) theory. Instead, following the recent literature we incorporate the source for the stress tensor as the modular parameter $\tau$ of the boundary torus. We refer to this choice as the $(\tau,\bar{\tau})$ formalism. We stress that there are subtle technical differences in both approaches, and it is then worth addressing the problem in detail.  

%%%%%%%%%%%%%%%%%%%%%%%%%%%%%%%%%%%%%%%%%%%%%%%
\subsection{Analytic continuation: Euclidean variational principle and solutions}
%%%%%%%%%%%%%%%%%%%%%%%%%%%%%%%%%%%%%%%%%%%%%%%
Let us start by defining the Euclidean section of the solutions. First, the Euclidean time direction is compactified and the topology of $M$ becomes that of a solid torus. We then introduce complex coordinates $(z,\bar{z})$ by analytically continuing the light-cone directions as $x^{+} \to z\,$, $x^{-}\to -\bar{z}\,$. The coordinate $z$ is identified as
\begin{equation}\label{identifications}
z \simeq z + 2\pi \simeq z + 2\pi \tau\,,
\end{equation}

\noindent where $\tau$ is the modular parameter of the boundary torus, with area $\text{Vol}(\partial M) = 4\pi^{2}\text{Im}(\tau)\,$. For the BTZ solution, for example, the absence of a coordinate singularity at the horizon requires
\begin{equation}\label{modular parameter}
\tau = \frac{i\beta}{2\pi}\left(1+\Omega\right)\,,\quad \bar{\tau} =  \frac{i\beta}{2\pi}\left(-1+\Omega\right)
\end{equation}

\noindent where the black hole temperature is $T=1/\beta$ and $\Omega$ is the angular velocity of the horizon ($\Omega$ and the angular momentum $J$ should be continued to purely imaginary values in order for the Euclidean section to be real). In the saddle-point approximation, valid for large temperatures and large central charges, the CFT partition function is obtained from the Euclidean on-shell action as
\begin{equation}
\ln Z = -I^{(E)}_{os} =\left. -\left(I^{(E)}_{CS} + I^{(E)}_{Bdy}\right)\right|_{os}\,,
\end{equation}

\noindent where
\begin{equation}\label{Euclidean Chern-Simons action}
I^{(E)}_{CS} =  \frac{ik_{cs}}{4\pi}\int_{M} \mbox{Tr}\Bigl[CS(A) - CS(\bar{A})\Bigr].
\end{equation}

Following  \cite{Gutperle:2011kf,Ammon:2011nk} we define smooth black hole solutions to be those for which the holonomy of the connection around the contractible cycle of the constant-$\rho$ torus is trivial, which is achieved if the holonomy belongs to the center of the gauge group \cite{Castro:2011fm,Castro:2011iw}. In the absence of rotation $\tau = -\bar{\tau} = i\beta/(2\pi)\,$ and the smoothness conditions boil down to the requirement that the holonomy around the Euclidean time circle is trivial. In the general case, the holonomies associated with the identification $z  \simeq z + 2\pi \tau$ are
\begin{align}\label{thermal holonomies}
\mbox{Hol}_{\tau,\bar{\tau}}(A) 
=b^{-1}e^{h}\,b\,,\qquad
\mbox{Hol}_{\tau,\bar{\tau}}(\bar{A}) 
=
b\, e^{\bar{h}}b^{-1}\, ,
\end{align}
\noindent where the matrices $h$ and $\bar{h}$ are defined as
\begin{equation}
h = 2\pi\left(\tau a_{z} +\bar{\tau}a_{\bar{z}}\right)\,,\qquad \bar{h} = 2\pi\left(\tau \bar{a}_{z} + \bar{\tau}\bar{a}_{\bar{z}}\right).
\end{equation}

\noindent The trivial holonomy requirement is then a restriction on the eigenvalues of $h$, $\bar{h}\,$. In general, the smoothness conditions will have multiple solutions, giving rise to different thermodynamic branches. For the $N=3$ black hole of \cite{Gutperle:2011kf} these possibilities were first studied in \cite{David:2012iu}. Here we focus in the so-called ``BTZ branch", defined by the requirement that the we recover the BTZ solution when all the higher spin sources and charges are turned off. A convenient way to encode the smoothness conditions in the BTZ branch is
\begin{equation}\label{smoothness conditions}
\text{spec}\Bigl(2\pi\left(\tau a_{z} +\bar{\tau}a_{\bar{z}}\right)\Bigr) = \text{spec}\left(2\pi i \Lambda^{0}\right) \quad \Rightarrow \quad \tau a_{z} +\bar{\tau}a_{\bar{z}} = u^{-1}\left(i\Lambda^{0}\right)u
\end{equation}

\noindent for some matrix $u\,$. The other thermodynamic branches emerge because of the possibility of reshuffling the eigenvalues of $h$ (and $\bar{h}$) while keeping the holonomy trivial, or because solutions exist for which the holonomy is conjugate to a different element of the center
of the gauge group. In this light, ``phase transitions" occur for values of the charges where different eigenvalues cross (i.e. become degenerate).

Let us now examine the Euclidean variational principle. Extra care must be exercised when computing the variation of the Chern-Simons action in the $(\bar{\tau},\tau)$ formalism: since the modular parameter of the torus is varying, we must write down \eqref{CS action variation} in coordinates with fixed periodicity before rewriting the result in terms of $(z,\bar{z})\,$. To this end, it is useful to consider the following change of coordinates \cite{Kraus:2006wn}:
\begin{equation}
z = \frac{1-i\tau}{2}w + \frac{1+i\tau}{2}\bar{w}\,
%,\qquad \bar{z} = \frac{1-i\bar{\tau}}{2}w + \frac{1 + i\bar{\tau}}{2}\bar{w}\,,
\end{equation}

\noindent so that the identifications now become
\begin{equation}
w \simeq w + 2\pi \simeq w + 2\pi i\,.
\end{equation}

\noindent Notice that the modular parameter now appears explicitly in the boundary metric:
\begin{equation}
ds_{(0)}^{2} = dzd\bar{z} = \left|\left(\frac{1-i\tau}{2}\right)dw + \left(\frac{1+i\tau}{2}\right)d\bar{w}\right|^{2}\,.
\end{equation}

\noindent Similarly, the components of the connection
%are related by 
%\begin{align}\label{connection in w}
%a_{w} 
%={}&
% \left(\frac{1-i\tau}{2}\right)a_{z} + \left(\frac{1-i\bar{\tau}}{2}\right)a_{\bar{z}}
%\\
%a_{\bar{w}}
% ={}&
% \left(\frac{1+i\tau}{2}\right)a_{z} + \left(\frac{1 + i\bar{\tau}}{2}\right)a_{\bar{z}}\, ,
%\end{align}
%
%\noindent which implies that the components 
corresponding to the non-contractible and contractible cycles of the torus are given respectively by 
\begin{equation}
a_{z} + a_{\bar{z}} = a_{w} + a_{\bar{w}}\,,\qquad \tau a_{z} + \bar{\tau}a_{\bar{z}} = i\left(a_{w}-a_{\bar{w}}\right),
\end{equation}

\noindent and the volume elements are related by
\begin{equation}\label{volume elements}
i\,dw\wedge d\bar{w}=\frac{2}{(\bar{\tau}-\tau)}dz\wedge d\bar{z} = i\frac{dz\wedge d\bar{z} }{\text{Im}\left(\tau\right)} = 2\frac{d^{2}z}{\text{Im}(\tau)}\,,
\end{equation}

\noindent where $d^{2}z$ is the standard measure on the Euclidean plane. Writing the on-shell variation of the Euclidean Chern-Simons action \eqref{Euclidean Chern-Simons action} in the $(w,\bar{w})$ coordinates we find
\begin{align}
\left. \delta I^{(E)}_{CS}\right|_{os}  
={}&
- \frac{ik_{cs}}{4\pi}\int_{\partial M}\text{Tr}\Bigl[a\wedge \delta a - \bar{a}\wedge \delta \bar{a}\Bigr]
\\
={}&
 - \frac{ik_{cs}}{4\pi}\int_{\partial M}dw\wedge d\bar{w}\,\text{Tr}\Bigl[a_{w}\delta a_{\bar{w}} - a_{\bar{w}}\delta a_{w}-\bar{a}_{w}\delta \bar{a}_{\bar{w}} + \bar{a}_{\bar{w}}\delta \bar{a}_{w}\Bigr].
\end{align}

\noindent  In order to pass back to the $(z,\bar{z})$ coordinates we note that the variations involve $\delta \tau$ and $\delta \bar{\tau}$ terms, e.g.
\begin{align}
\delta a_{w} 
={}&
 \left(\frac{1-i\tau}{2}\right)\delta a_{z} + \left(\frac{1-i\bar{\tau}}{2}\right)\delta a_{\bar{z}} -\frac{i}{2}
 \left(\delta \tau\, a_{z} + \delta \bar{\tau}\,a_{\bar{z}}\right).
\end{align}

\noindent Taking this and \eqref{volume elements} into account, the on-shell variation in the $(z,\bar{z})$ parameterization is
\begin{align}\label{CS action variation in z zbar}
\left. \delta I^{(E)}_{CS}\right|_{os} ={}&
 -i\pi k_{cs}\int_{\partial M} \frac{d^{2}z}{4\pi^{2}\text{Im}\left(\tau\right)}\text{Tr}\Bigl[\left(\bar{\tau}-\tau\right)\left(a_{z}\delta a_{\bar{z}} - a_{\bar{z}}\delta a_{z}\right) + \left(a_{z}+a_{\bar{z}}\right)\left(\delta \tau\, a_{z} + \delta \bar{\tau}\,a_{\bar{z}}\right)\Bigr] 
\nonumber\\
 &
 +i\pi k_{cs}\int_{\partial M} \frac{d^{2}z}{4\pi^{2}\text{Im}\left(\tau\right)}\text{Tr}\Bigl[\left(\bar{\tau}-\tau\right)\left(\bar{a}_{z}\delta \bar{a}_{\bar{z}} - \bar{a}_{\bar{z}}\delta \bar{a}_{z}\right) + \left(\bar{a}_{z}+\bar{a}_{\bar{z}}\right)\left(\delta \tau\, \bar{a}_{z} + \delta \bar{\tau}\,\bar{a}_{\bar{z}}\right)\Bigr] .
\end{align}

Next, consider the Euclidean continuation of the boundary term \eqref{Lorentzian boundary term},
\begin{equation}\label{Euclidean boundary term}
I^{(E)}_{Bdy} =-\frac{k_{cs}}{2\pi} \int_{\partial M} d^{2}z\, \text{Tr}\left[\left(a_{z} -2 \Lambda^{+}\right)a_{\bar{z}}\right] 
-\frac{k_{cs}}{2\pi}\int_{\partial M} d^{2}z\, \text{Tr}\left[\left(\bar{a}_{\bar{z}} -2 \Lambda^{-}\right)\bar{a}_{z} \right],
\end{equation}

\noindent with variation\footnote{Notice that we must keep fixed the invariant measure \eqref{volume elements}  in this variation.}
\begin{align}\label{boundary term variation in z zbar}
\delta I^{(E)}_{Bdy} ={}&
 -i \frac{k_{cs}}{4\pi} \int_{\partial M} \frac{d^{2}z}{\text{Im}(\tau)} \,\delta\Bigl(\left(\bar{\tau}-\tau\right) \text{Tr}\left[\left(a_{z} -2 \Lambda^{+}\right)a_{\bar{z}}\right]\Bigr) 
\nonumber\\
&
-i\frac{k_{cs}}{4\pi}\int_{\partial M} \frac{d^{2}z}{\text{Im}(\tau)} \,\delta\Bigl(\left(\bar{\tau}-\tau\right) \text{Tr}\left[\left(\bar{a}_{\bar{z}} -2 \Lambda^{-}\right)\bar{a}_{z} \right]\Bigr).
\end{align}

\noindent Combining \eqref{CS action variation in z zbar} and \eqref{boundary term variation in z zbar} we arrive at
\begin{align}\label{Euclidean variation 1}
\delta I^{(E)}_{os} =-2\pi i k_{cs}\int_{\partial M}\frac{d^{2}z}{4\pi^{2}\text{Im}(\tau)}\text{Tr}&\left[ \left(a_{z}-\Lambda^{+}\right)\delta\Bigl(\left(\bar{\tau}-\tau\right)a_{\bar{z}}\Bigr) +\left(\frac{a_{z}^{2}}{2} + a_{z}a_{\bar{z}} -\frac{\bar{a}_{z}^{2}}{2}\right)\delta \tau 
\right.
\nonumber\\
&\left.
- \left(-\bar{a}_{\bar{z}}+\Lambda^{-}\right)\delta\Bigl(\left(\bar{\tau} - \tau\right)\bar{a}_{z}\Bigr) -\left( \frac{\bar{a}_{\bar{z}}^{2}}{2} + \bar{a}_{\bar{z}}\bar{a}_{z} -\frac{a_{\bar{z}}^{2}}{2}\right)\delta\bar{\tau}\right].
\end{align}

\noindent We then see that the variational principle provides a natural way to identify the structure of the Euclidean solutions, and in particular the definition of the sources and EVs. Much like in the Lorentzian case, we write the connections in Drinfeld-Sokolov form
\begin{align}\label{Euc DS connections}
a ={}&
 \Bigl(\Lambda^{+} + Q\Bigr)dz -\Bigl(M + \ldots \Bigr)d\bar{z}
 \\
\bar{a} ={}&
 \Bigl(\Lambda^{-} - \bar{Q}\Bigr)d\bar{z} + \Bigl(\bar{M} + \ldots \Bigr)dz
\label{Euc DS connections 2}
\end{align}

\noindent with $\left[\Lambda^{-},Q\right] = \left[\Lambda^{+},M\right]=0$ (and similarly for the $\bar{Q}$, $\bar{M}$) with two main differences: $M$ and $\bar{M}$ will not contain $\mu_{2}$, $\bar{\mu}_{2}$ which are already explicitly present in the formulas as $\bar{\tau}$, $\tau$, and the definition of the higher spin chemical potentials is now
\begin{align}\label{Euclidean normalizations}
\text{Tr}\left[\left(a_{z}-\Lambda^{+}\right)\left(\bar{\tau}-\tau\right)a_{\bar{z}}\right] ={}&
 \text{Tr}\left[Q\left(\bar{\tau}-\tau\right)a_{\bar{z}}\right] = \sum_{j=3}^{N}\mu_{j}Q_{j}
 \\
 \text{Tr}\left[\left(-\bar{a}_{\bar{z}}+\Lambda^{-}\right)\left(\bar{\tau}-\tau\right)\bar{a}_{z}\right]  ={}&
   \text{Tr}\left[\bar{Q}\left(\bar{\tau}-\tau\right)\bar{a}_{z}\right] = \sum_{j=3}^{N}\bar{\mu}_{j}\bar{Q}_{j}\,.
\label{Euclidean normalizations 1b}
\end{align}

\noindent For simplicity we have written \eqref{Euclidean normalizations}-\eqref{Euclidean normalizations 1b} as appropriate to the principal embedding. Following the Lorentzian signature discussion (e.g. \eqref{mu Q terms diag}-\eqref{mu Q terms 2 diag}) the extension to other embeddings is straightforward. 

In addition, we see that the variation \eqref{Euclidean variation 1} allows us to identify the zero modes of the stress tensor as
\begin{equation}\label{stress tensor formula}
T =\text{Tr}\left[ \frac{a_{z}^{2}}{2} + a_{z}a_{\bar{z}} -\frac{\bar{a}_{z}^{2}}{2}\right] \,,\qquad \bar{T} =\text{Tr}\left[  \frac{\bar{a}_{\bar{z}}^{2}}{2} + \bar{a}_{\bar{z}}\bar{a}_{z} -\frac{a_{\bar{z}}^{2}}{2}\right].
\end{equation}

\noindent We note the rather unusual feature that, due to the nonlinear character of the asymptotic symmetry algebra and the relaxation of the boundary conditions \eqref{AAdS3 bcs} to allow for irrelevant deformations, the left-moving component of the stress tensor receives contributions from the right-moving connection, and vice versa.

 Having identified the sources and EVs as indicated above, \eqref{Euclidean variation 1} becomes
\begin{align}\label{Euclidean variation}
\delta \ln Z = -\delta I^{(E)}_{os} = 2\pi i k_{cs}\int_{\partial M}\frac{d^{2}z}{4\pi^{2}\text{Im}(\tau)}\Biggl(T\delta\tau -\bar{T}\delta\bar{\tau}+ \sum_{j=3}^{N}\left(Q_{j}\delta\mu_{j}-\bar{Q}_{j}\delta\bar{\mu}_{j}\right) \Biggr),
\end{align}

\noindent as expected for the variation of the partition function in the principal embedding. As in the Lorentzian case, we stress that this result holds even for solutions with non-constant sources and charges, because the equations \eqref{Euc DS connections}-\eqref{Euclidean normalizations 1b} defining the lowest/highest weight structure of the solutions remain valid. In the special case of constant connections $a$, $\bar{a}$, using $\text{Vol}(\partial M) = 4\pi^{2}\text{Im}(\tau)\,$ we find that the above result (principal embedding) reduces to
\begin{equation}\label{Euclidean variation constant case}
\text{constant $a$, $\bar{a}$:} \quad \delta \ln Z  = 2\pi i k_{cs}\Biggl(T\delta\tau -\bar{T}\delta\bar{\tau}+ \sum_{j=3}^{N}\left(Q_{j}\delta\mu_{j}-\bar{Q}_{j}\delta\bar{\mu}_{j}\right) \Biggr).
\end{equation}

%%%%%%%%%%%%%%%%%%%%%%%%%%%%%%%%%%%%%%%%%%%%%%%
\subsection{Entropy and free energy}\label{subsec:entropy}
%%%%%%%%%%%%%%%%%%%%%%%%%%%%%%%%%%%%%%%%%%%%%%%
For solutions consisting of constant connections $a$, $\bar{a}$, we can explicitly evaluate the Euclidean action \eqref{Euclidean Chern-Simons action} plus the boundary term \eqref{Euclidean boundary term} on-shell,\footnote{See \cite{Banados:2012ue} for subtleties associated with the on-shell evaluation of the bulk piece.} obtaining the free energy $F$ as
\begin{align}\label{canonical free energy}
-\beta F =
 \ln Z
  ={}&
   -2\pi i k_{cs}\text{Tr}\biggl[\tau\left(\frac{a_{z}^{2}}{2} + a_{z}a_{\bar{z}} -\frac{\bar{a}_{z}^{2}}{2}\right) -\bar{\tau}\left( \frac{\bar{a}_{\bar{z}}^{2}}{2} + \bar{a}_{\bar{z}}\bar{a}_{z} -\frac{a_{\bar{z}}^{2}}{2}\right)\nonumber\\
   {}&\hphantom{-2\pi i k_{cs}\text{Tr}\Bigl[ a}
     + \left(\bar{\tau}-\tau\right)\left(\Lambda^+ a_{\bar{z}}+\Lambda^{-} \,\bar{a}_{z}\right)\biggr].
\end{align}

\noindent From \eqref{Euclidean variation 1} we can immediately read-off the term that performs the Legendre transform from the free energy (a function of the sources) to the entropy (a function of the charges), namely
\begin{align}
S =
\ln  Z  -2\pi i k_{cs}\text{Tr}&\left[ \left(\bar{\tau}-\tau\right)\left(a_{z}-\Lambda^{+}\right)a_{\bar{z}} +\tau\left(\frac{a_{z}^{2}}{2} + a_{z}a_{\bar{z}} -\frac{\bar{a}_{z}^{2}}{2}\right)  
\right.
\nonumber\\
&\quad \left.
- \left(\bar{\tau} - \tau\right)\left(-\bar{a}_{\bar{z}}+\Lambda^{-}\right)\bar{a}_{z}-\bar{\tau}\left( \frac{\bar{a}_{\bar{z}}^{2}}{2} + \bar{a}_{\bar{z}}\bar{a}_{z} -\frac{a_{\bar{z}}^{2}}{2}\right)\right].
\end{align}

\noindent Plugging \eqref{canonical free energy} into this equation we conclude that the higher spin black hole entropy in the $(\bar{\tau},\tau)$ formalism is given by
\begin{equation}\label{our entropy formula}
S = -2\pi i k_{cs}\,\text{Tr}\Bigl[\left(a_{z}+a_{\bar{z}}\right)\left(\tau a_{z} +\bar{\tau}a_{\bar{z}}\right)-\left(\bar{a}_{z}+\bar{a}_{\bar{z}}\right)\left(\tau\bar{a}_{z} + \bar{\tau}\bar{a}_{\bar{z}}\right)\Bigr],
\end{equation}

\noindent with $k_{cs}$ given by \eqref{kcs}. As a first check of our results, we will show that for smooth solutions (i.e. when the holonomy conditions are satisfied), the variation of the entropy has the correct form. To this end we first rewrite $S$ as
 \begin{equation}\label{entropy in XY}
S = -2\pi i k_{cs}\text{Tr}\Bigl[XY-\bar{X}\bar{Y}\Bigr],
\end{equation}

\noindent where $X\equiv \left(a_{z}+a_{\bar{z}}\right)$, $Y\equiv \left(\tau a_{z} +\bar{\tau}a_{\bar{z}}\right)$ and similarly for the barred quantities. From \eqref{smoothness conditions} we know that the smoothness condition in the BTZ branch implies that the component of the connection along the contractible cycle has the form
\begin{equation}\label{holonomy conditions v2}
2\pi Y = u^{-1} \left(2\pi i \Lambda^{0}\right)u\,.
\end{equation}

\noindent Hence, under a variation with infinitesimal parameter $\epsilon$ we have 
\begin{equation}
\delta Y = \left[Y,\epsilon\right].
\end{equation}

\noindent Using this fact together with $\left[X,Y\right]=0$ (from the equations of motion) it is immediate to prove that, for smooth constant connections, 
\begin{equation}
\text{Tr}\left[Y\delta Y\right] =\text{Tr}\left[X\delta Y\right] =0\,.
\end{equation}

\noindent Using these relations we easily obtain
\begin{align}\label{variation entropy}
\delta S 
={}&
  -2\pi i k_{cs}\text{Tr}\Bigl[\delta X\,Y-\delta \bar{X}\,\bar{Y}\Bigr]
  \nonumber\\
  ={}&
    -2\pi i k_{cs}\text{Tr}\Biggl[ \tau\, \delta\left( \frac{a_{z}^{2}}{2} + a_{z}a_{\bar{z}} -\frac{\bar{a}_{z}^{2}}{2} \right)-\bar{\tau}\,\delta \left( \frac{\bar{a}_{\bar{z}}^{2}}{2} + \bar{a}_{\bar{z}}\bar{a}_{z} -\frac{a_{\bar{z}}^{2}}{2}\right)
    \nonumber\\
    &\hphantom{ -2\pi i k_{cs}\text{Tr}\Biggl[}
    +\left(\bar{\tau}-\tau\right)a_{\bar{z}} \,\delta \left(a_{z}-\Lambda^{+}\right) - \left(\bar{\tau}-\tau\right)\bar{a}_{z}\,\delta\left(-\bar{a}_{\bar{z}}+\Lambda^{-}\right)\Biggr]
    \nonumber\\
    ={}&
     -2\pi i k_{cs}\biggl[ \tau\, \delta T-\bar{\tau}\,\delta \bar{T}+\sum_{j=3}^{N}\left(\mu_{j}\delta Q_{j}-\bar{\mu}_{j}\delta\bar{Q}_{j}\right)\biggr]
\end{align}

\noindent which is precisely the Legendre transform of \eqref{Euclidean variation constant case}, as expected. 

In order to give an explicit expression for the entropy and free energy in terms of the charges and chemical potentials, we now return to the problem of specifying sources and EVs in our solution. As we already mentioned in section \ref{Lorentzian bhs}, in general the conditions \eqref{mu Q terms}-\eqref{mu Q terms 2} (Lorentzian signature) or \eqref{Euclidean normalizations}-\eqref{Euclidean normalizations 1b} (Euclidean signature) do not completely fix the definition of the chemical potentials. For constant connections in the principal embedding, the remaining freedom can be used to require in addition\footnote{The analogous requirement for Lorentzian solutions in the principal embedding is $\text{Tr}\Bigl[\Lambda^{+}a_{-}\Bigr] = \sum_{j =2}^{N}(j-1)\mu_{j}Q_{j}\,$ and $\text{Tr}\Bigl[-\Lambda^{-}\bar{a}_{+}\Bigr] = \sum_{j =2}^{N}(j-1)\bar{\mu}_{j}\bar{Q}_{j}\,$.}
\begin{align}\label{Euclidean normalizations 2a}
\text{Tr}\Bigl[\Lambda^{+}\left(\bar{\tau}-\tau\right)a_{\bar{z}}\Bigr] ={}& \sum_{j =3}^{N}(j-1)\mu_{j}Q_{j}
\\
\text{Tr}\Bigl[-\Lambda^{-}\left(\bar{\tau}-\tau\right)\bar{a}_{z}\Bigr] ={}& \sum_{j =3}^{N}(j-1)\bar{\mu}_{j}\bar{Q}_{j}\,.
\label{Euclidean normalizations 2b}
\end{align}

\noindent A similar expression applies to non-principal embeddings, with $j$ replaced by the conformal weight of the corresponding operators, and the sum running over the fields present in the spectrum. As a concrete example, let us again consider the constant solution in the $N=3$ theory. According to our definitions one finds\footnote{For constant connections, Cayley-Hamilton's theorem ensures that one can always write $a_{\bar{z}}$ as a linear combination of the matrices $m_{(n)} \equiv a_{z}^{n} - \frac{\mathds{1}}{N}\text{Tr}\left[a_{z}^{n}\right]$. However, for $N>3$ the coefficient of $m_{(n)}$ in this expansion is not necessarily $\mu_{n}/(\bar{\tau}-\tau)\,$; the precise coefficient is fixed by requiring \eqref{Euclidean normalizations 2a}-\eqref{Euclidean normalizations 2b} to be satisfied.}
\begin{equation}
a_{z} = \left(
\begin{array}{ccc}
0 & \frac{Q_{2}}{2} & Q_{3} \\ 
1 & 0 & \frac{Q_{2}}{2}\\ 
0 & 1 & 0
\end{array} \right)\,,\qquad a_{\bar{z}} = \frac{\mu_{3}}{\bar{\tau}-\tau}\left(a_{z}^{2} - \frac{\mathds{1}}{3}\text{Tr}\left[a_{z}^{2}\right]\right) =  \frac{\mu_{3}}{\bar{\tau}-\tau}\left(\begin{array}{ccc}
-\frac{Q_{2}}{6}& Q_{3} & \frac{Q_{2}^{2}}{4} \\ 
0 & \frac{Q_{2}}{3}& Q_{3} \\ 
1& 0 & -\frac{Q_{2}}{6}
\end{array} \right),
\end{equation}

\noindent where $\mathds{1}$ denotes the identity matrix. Similar expressions hold for the barred connection. We stress that according to \eqref{stress tensor formula} the energy is not just $Q_{2}$, but rather
\begin{equation}
T = Q_{2} +3 \left(\frac{ \mu_{3}Q_{3}}{\bar{\tau}-\tau} \right)- \frac{1}{3}\left(\frac{\bar{\mu}_{3}\bar{Q}_{2}}{\bar{\tau}-\tau}\right)^{2}\,,
\end{equation}

\noindent with a similar expression for $\bar{T}\,$.

Going back to the general case, direct evaluation of \eqref{our entropy formula} and \eqref{canonical free energy} using \eqref{Euclidean normalizations}-\eqref{Euclidean normalizations 1b} and \eqref{Euclidean normalizations 2a}-\eqref{Euclidean normalizations 2b} yields (in the principal embedding)
\begin{equation}
S =  -2\pi i k_{cs}\Biggl[2\tau T -2\bar{\tau}\bar{T} + \sum_{j=3}^{N}j\left(\mu_{j}Q_{j} - \bar{\mu}_{j}\bar{Q}_{j}\right)\Biggr]
\end{equation}

\noindent and
\begin{equation}
-\beta F = \ln Z =  -2\pi i k_{cs}\Biggl[\tau T -\bar{\tau}\bar{T} + \sum_{j=3}^{N}\left(j-1\right)\left(\mu_{j}Q_{j} - \bar{\mu}_{j}\bar{Q}_{j}\right)\Biggr],
\end{equation}

\noindent which have the expected form (see e.g. \cite{Banados:2012ue}). A similar expression is obtained in non-principal embeddings, with $j$ replaced by the conformal weight of the corresponding operator, and the sum running over the appropriate spectrum. 

Finally, we notice that our entropy formula \eqref{our entropy formula} can be rewritten in a remarkably simple form. Since for constant connections the matrices $X$ and $Y$ commute by the equations of motion, they can be simultaneously diagonalized. Hence, when the holonomy conditions \eqref{holonomy conditions v2} are obeyed, \eqref{entropy in XY} evaluates to
\begin{equation}\label{our entropy formula v2}
S = 2\pi  k_{cs}\text{Tr}\Bigl[\left( \lambda_{\varphi}- \bar{\lambda}_{\varphi}\right)\Lambda^{0}\Bigr],
\end{equation}

\noindent where $\lambda_{\varphi}$ and $\bar{\lambda}_{\varphi}$ are the diagonal matrices whose entries contain the eigenvalues of $X$ and $\bar{X}$, i.e.
\begin{equation}
\lambda_{\varphi} \equiv \text{Eigen}\left(a_{z}+a_{\bar{z}}\right) = \text{Eigen}\left(a_{\varphi}\right)\,,\qquad \bar{\lambda}_{\varphi} \equiv \text{Eigen}\left(\bar{a}_{z}+\bar{a}_{\bar{z}}\right) = \text{Eigen}\left(\bar{a}_{\varphi}\right)\,.
\end{equation}

\noindent Therefore, we have reduced the problem of evaluating the higher spin black hole entropy to the diagonalization of the component of the connection along the non-contractible cycle of the Euclidean torus, namely the black hole horizon. Indeed, for (constant) smooth solutions all the gauge-invariant information is contained in the horizon holonomy, and it is satisfying to see explicitly how the black hole entropy depends on this information only. 

As a side remark, in computing the entropy using \eqref{our entropy formula v2}, one might wonder whether the ordering of the eigenvalues is important. At the BTZ point there is a particular ordering where the matrix of eigenvalues is proportional to $\Lambda^0\,$. As one turns on higher spin sources, one should adiabatically change the eigenvalues, but not their ordering. This prescription works fine in a neighborhood of the BTZ black hole, but could become ambiguous once eigenvalues start to cross, and in particular multiple phases could appear. The expression \eqref{our entropy formula} for the entropy in terms of the gauge fields is unambiguous however, and one could always revert back to that one in order to determine the right ordering of the eigenvalues.

%%%%%%%%%%%%%%%%%%%%%%%%%%%%%%%%%%%%%%%%%%%%%%%%%%%%%%%%%%%%%%%%%%%%%%%%%%%%%%%%
\section{Discussion}\label{sec: conclusions}
%%%%%%%%%%%%%%%%%%%%%%%%%%%%%%%%%%%%%%%%%%%%%%%%%%%%%%%%%%%%%%%%%%%%%%%%%%%%%%%%
Even though our formalism applies to general black holes solutions and arbitrary $N$, it is instructive to apply our formulas in a few explicit examples in the $N=2$ and $N=3$ theories.  When appropriate, we will comment on the differences with respect to the results obtained with other approaches.

%%%%%%%%%%%%%%%%%%%%%%%%%%%%%%%%%%%%%%%%%%%%%%%
\subsection{$N=2$ and $N=3$ examples}
%%%%%%%%%%%%%%%%%%%%%%%%%%%%%%%%%%%%%%%%%%%%%%%
Let us first consider the simplest solution, namely the BTZ solution in the $N=2$ (pure gravity) theory. In the conventions of \cite{Ammon:2012wc} one finds $k_{cs}=k$ for the fundamental representation, and the Euclidean connections are given by
\begin{equation}
a = \left(\begin{array}{cc}
0 & 2\pi \mathcal{L}/k \\ 
1& 0
\end{array} \right)dz \,,\qquad \bar{a} = 
-\left(\begin{array}{cc}
0 & 1 \\ 
2\pi \bar{\mathcal{L}}/k & 0
\end{array} \right)d\bar{z}\,.
\end{equation}

\noindent Evaluating \eqref{our entropy formula} we find the well-known result
\begin{equation}\label{BTZ entropy}
S_{BTZ} = -4\pi^{2}i\left(2\tau \mathcal{L} - 2\bar{\tau}\mathcal{L}\right) = 2\pi\left(\sqrt{2\pi k\mathcal{L}} + \sqrt{2\pi k \bar{\mathcal{L}}}\,\right),
\end{equation}

\noindent where in the second equality we used the fact that the smoothness conditions in this case amount to \cite{Gutperle:2011kf}
\begin{equation}
\tau = \frac{ik}{2}\frac{1}{\sqrt{2\pi k\mathcal{L}}}\,,\qquad \bar{\tau} = -\frac{ik}{2}\frac{1}{\sqrt{2\pi k \bar{\mathcal{L}}}}\,.
\end{equation}

\noindent Equivalently, we can evaluate the entropy using \eqref{our entropy formula v2}. In this case we have
\begin{align}
\lambda_{\varphi} ={}&
 \text{Eigen}\left(a_{\varphi}\right) =\text{diag}\left(\sqrt{\frac{2\pi \mathcal{L}}{k}},-\sqrt{\frac{2\pi \mathcal{L}}{k}}\right)\,,
 \\
  \bar{\lambda}_{\varphi} ={}&
   \text{Eigen}\left(\bar{a}_{\varphi}\right)=\text{diag}\left(-\sqrt{\frac{2\pi \bar{\mathcal{L}}}{k}},\sqrt{\frac{2\pi \bar{\mathcal{L}}}{k}}\right)\,.
\end{align}

\noindent Using $\Lambda^{0} = \text{diag}\left(1/2,-1/2\right)$ and \eqref{our entropy formula v2} we find \eqref{BTZ entropy} again.

Similarly, we can consider the static $N=3$  black hole in the diagonal embedding \cite{Castro:2011fm},
\begin{align}\label{diagonal embedding connection}
a
 ={}&
\Bigl(W_{2}+w W_{-2}-qW_{0}\Bigr)dz - \frac{\eta}{2} W_0 d\bar{z} \, ,
\\
\bar{a} 
={}&
 -\Bigl(W_{-2}+ w W_{2}-qW_{0}\Bigr)d\bar{z} +\frac{\eta}{2} W_0 dz\, .
 \label{diagonal embedding connection 2}
\end{align}

\noindent In the diagonal embedding one has $\Lambda^{\pm} \sim W_{\pm 2}\,$, so the connections are written in Drinfeld-Sokolov form. This solution corresponds to a BTZ black hole carrying $U(1)$ charges. In the conventions of \cite{Castro:2011fm,Ammon:2012wc}, in the diagonal embedding one has $\text{Tr}\left[\Lambda^0 \Lambda^0\right] =1/2$ and therefore $k_{cs} = k\,$. Applying our formula \eqref{our entropy formula} and using the smoothness conditions $\eta = 2q$ and $\tau = i/(8\sqrt{\omega})$ \cite{Castro:2011fm} we obtain
\begin{equation}
S = -\frac{8\pi i}{3}k\tau\left(4q^{2} + 48w-\eta^{2}\right) = 16\pi k\sqrt{w}\,,
\end{equation}

\noindent which agrees with the result in \cite{Castro:2011fm}. We can also use our formula \eqref{our entropy formula v2} directly. For this solution, in the BTZ branch we find
\begin{align}
\lambda_{\varphi} ={}&
 \text{Eigen}\left(a_{\varphi}\right) =\text{diag}\left(4\sqrt{w} - \frac{1}{3}\left(\eta + 2q\right),\frac{2}{3}\left(\eta+2q\right),-4\sqrt{w} - \frac{1}{3}\left(\eta + 2q\right)\right)\,,
 \\
  \bar{\lambda}_{\varphi} ={}& -\lambda_{\varphi} \,.
\end{align}

\noindent In the diagonal embedding one has $\Lambda^{0} =\text{diag}\left(1/2,0,-1/2\right)$, and therefore \eqref{our entropy formula v2} yields $S=16\pi k\sqrt{w}$ as before.

We now turn our attention to the principal embedding $N=3$ black hole solutions of \cite{Gutperle:2011kf,Ammon:2011nk}, which exhibits some interesting features that deserve special attention. In the conventions of \cite{Ammon:2012wc}, the connections read
\begin{align}\label{highest weight gauge connection Gutperle Kraus}
a
={}&
\biggl( L_{1}-\frac{2\pi \mathcal{L}}{k}\, L_{-1}-\frac{\pi \mathcal{W}}{2k}\,W_{-2}\biggr) dz -\mu \Biggl(W_{2}+\frac{4\pi \mathcal{W}}{k}\,L_{-1}+\left( \frac{2\pi \mathcal{L}}{k}\right)^{2}\, W_{-2}-\frac{4\pi \mathcal{L}}{k}\,W_0 \Biggr) d\bar{z}\, ,
\\
\nonumber
\bar{a}
={}&
\biggl( L_{-1}-\frac{2\pi \bar{\mathcal{L}}}{k}\, L_{1}-\frac{\pi \bar{\mathcal{W}}}{2k}\,W_{2}\biggr) d\bar{z} -\bar{\mu} \Biggl(W_{-2}+\frac{4\pi \bar{\mathcal{W}}}{k}\,L_{1}+\left( \frac{2\pi \bar{\mathcal{L}}}{k}\right)^{2}\, W_{2}-\frac{4\pi \bar{\mathcal{L}}}{k}\,W_0 \Biggr) dz\,,
%\label{highest weight gauge connection Gutperle Kraus 2}
\end{align}

\noindent where $\mathcal{W}$ and $\alpha=\mu\bar{\tau}$ are, respectively, the spin-3 charge and source, with similar expressions in the other chiral sector. Since the above connections are written in Drinfeld-Sokolov form, our general formulas for the energy and entropy apply. We note that in the normalization of the generators in \cite{Ammon:2012wc} one has $\text{Tr}\left[\Lambda^0 \Lambda^0\right]=\text{Tr}\left[L_0 L_0\right]= 2$ and therefore the Chern-Simons level \eqref{kcs} is $k_{cs} = k/4\,$. The energy is easily obtained from \eqref{stress tensor formula} as
\begin{align}
T ={}&
 \frac{2\pi}{k_{cs}}\left(\mathcal{L} + 3\mu \mathcal{W} - \frac{8\pi}{3k_{cs}}\bar{\mu}^{2}\bar{\mathcal{L}}^{2} \right),
\\
\bar{T} ={}&
\frac{2\pi}{k_{cs}}\left(\bar{\mathcal{L}} + 3\bar{\mu} \bar{\mathcal{W}} - \frac{8\pi}{3k_{cs}}\mu^{2}\mathcal{L}^{2} \right).
\end{align}

\noindent Similar $\mu$-dependent terms appear in the black hole energy (mass) computed via canonical methods \cite{Perez:2012cf}. In the latter approach the precise expression for the integrated charge depends on the choice of boundary conditions. In our treatment the boundary conditions were specified in the variational principle, and one could in principle map between the two formalisms. We note however that the left-moving energy receives a non-linear contribution from the right-movers, and vice-versa, a peculiar feature that obscures the interpretation of these quantities in CFT language.

Evaluating our general entropy formula \eqref{our entropy formula} in the solution \eqref{highest weight gauge connection Gutperle Kraus} we obtain
\begin{equation}\label{rotating GK entropy}
S = -4\pi^{2}i\left(2\tau \mathcal{L} - 2\bar{\tau}\bar{\mathcal{L}} + \frac{16\pi}{3k_{cs}}\bar{\tau}\mu^{2}\mathcal{L}^{2}- \frac{16\pi}{3k_{cs}}\tau\bar{\mu}^{2}\bar{\mathcal{L}}^{2} + 3\left(\tau+\bar{\tau}\right)\left(\mu\mathcal{W}-\bar{\mu}\bar{\mathcal{W}}\right)\right).
\end{equation}

\noindent In the non-rotating limit $\bar{\tau}=-\tau$, $\bar{\mathcal{L}}=\mathcal{L}$, $\bar{\mathcal{W}}=-\mathcal{W}$, and the holonomy conditions can be solved explicitly: in the conventions of  \cite{Gutperle:2011kf,Ammon:2011nk} one finds
\begin{equation}\label{solution smoothness GK}
\mathcal{W} = \frac{4(C-1)}{C^{3/2}}\mathcal{L}\sqrt{\frac{2\pi \mathcal{L}}{k}}\,,\qquad \mu = \frac{3\sqrt{C}}{4(2C-3)}\sqrt{\frac{k}{2\pi \mathcal{L}}}\,,\qquad \tau = \frac{i\left(2C-3\right)}{4\left(C-3\right)\sqrt{1-\frac{3}{4C}}}\sqrt{\frac{k}{2\pi \mathcal{L}}}\,,
\end{equation}

\noindent where $C > 3$ and $C =\infty$ at the BTZ point. In the non-rotating case \eqref{rotating GK entropy} then reduces to
\begin{align}\label{static spin 3 result}
S_{J=0} ={}&
 -8\pi^{2}i\left(2\tau \mathcal{L} - \frac{16\pi}{3k_{cs}}\tau\mu^{2}\mathcal{L}^{2} \right)
 \nonumber\\
  ={}&
   4\pi\sqrt{2\pi k \mathcal{L}}\left(1-\frac{3}{2C}\right)^{-1}\sqrt{1 -\frac{3}{4C}}\,\,.
\end{align}

\noindent Let us repeat the calculation using the simpler formula \eqref{our entropy formula v2}. In the non-rotating case, using \eqref{solution smoothness GK} we find the eigenvalues of $a_{\varphi}$ in the BTZ branch as
\begin{align}
\lambda_{\varphi} ={}&
% \text{Eigen}\left(a_{\varphi}\right) =2\sqrt{\frac{2\pi \mathcal{L}}{k}}\text{diag}\left(\frac{3 -C\left(2 - \sqrt{4C-3}\right)}{\sqrt{C}\left(2C-3\right)},\frac{2}{\sqrt{C}},\frac{3 -C\left(2 + \sqrt{4C-3}\right)}{\sqrt{C}\left(2C-3\right)}\right)\,,
 \text{Eigen}\left(a_{\varphi}\right) =2\sqrt{\frac{2\pi \mathcal{L}}{k}}\,\text{diag}\left(-\frac{1}{\sqrt{C}} + \frac{\sqrt{1 -\frac{3}{4C}}}{1-\frac{3}{2C}},\,\frac{2}{\sqrt{C}}\,,-\frac{1}{\sqrt{C}} -\frac{\sqrt{1 -\frac{3}{4C}}}{1-\frac{3}{2C}}\right)\,,
 \\
  \bar{\lambda}_{\varphi} ={}& -\lambda_{\varphi} \,.
\end{align}

\noindent In the principal embedding we have $\Lambda^{0}=L_{0}=\text{diag}\left(1,0,-1\right)$ and $k_{cs} = k/4\,$, so using \eqref{our entropy formula v2} we recover \eqref{static spin 3 result}.

Our expression \eqref{static spin 3 result} agrees with the result obtained  in \cite{Perez:2013xi} via canonical methods, and also with the (perturbative) result obtained in \cite{Campoleoni:2012hp} using Wald's formula. As a third independent check, one can show that this is the result obtained by taking the appropriate limit in entanglement entropy calculations in the higher spin theory \cite{deBoer:2013}. We note however that the above result for the entropy of the static spin-3 black hole differs from the one first obtained by Gutperle and Kraus in \cite{Gutperle:2011kf}, which we denote by $S_{\text{G-K}}\,$:
\begin{equation}\label{GK result}
S_{\text{G-K}}= 4\pi\sqrt{2\pi k \mathcal{L}}\,\sqrt{1 -\frac{3}{4C}} = \left(1-\frac{3}{2C}\right)S_{J=0}\,.
\end{equation}

\noindent The origin of the discrepancy is clear: the entropy is obtained by demanding compatibility with the first law of thermodynamics (existence of the partition function); hence, different definitions of the energy (i.e. the black hole mass) will produce different results for the entropy. While the definition used in \cite{Gutperle:2011kf,Ammon:2011nk} is natural from the point of view of the holomorphic structure of the dual CFT, our result is consistent with the canonical definitions of free energy and entropy that follow from the Euclidean variational principle with quite natural boundary conditions. Below we will further elaborate on this point.

%%%%%%%%%%%%%%%%%%%%%%%%%%%%%%%%%%%%%%%%%%%%%%%
\subsection{Comparison with the ``holomorphic'' formalism}
%%%%%%%%%%%%%%%%%%%%%%%%%%%%%%%%%%%%%%%%%%%%%%%
To better understand the relation between our result and that of \cite{Gutperle:2011kf,Ammon:2011nk}, we first observe that \eqref{GK result} can be obtained from an expression very similar to \eqref{our entropy formula}, namely
\begin{equation}\label{our holomorphic formula}
S_{\text{G-K}} = -2\pi i k_{cs}\,\text{Tr}\Bigl[a_{z}\left(\tau a_{z} +\bar{\tau}a_{\bar{z}}\right)-\bar{a}_{\bar{z}}\left(\tau\bar{a}_{z} + \bar{\tau}\bar{a}_{\bar{z}}\right)\Bigr].
\end{equation}

\noindent Repeating the reasoning that led us from \eqref{our entropy formula} to \eqref{our entropy formula v2}, in this case we can show that
\begin{equation}
S_{\text{G-K}} = 2\pi  k_{cs}\,\text{Tr}\Bigl[\left(\lambda_{z}-\bar{\lambda}_{\bar{z}}\right)\Lambda^{0}\Bigr],
\end{equation}

\noindent where
\begin{equation}
\lambda_{z} \equiv \text{Eigen}\left(a_{z}\right)\,,\qquad \bar{\lambda}_{\bar{z}} \equiv \text{Eigen}\left(\bar{a}_{\bar{z}}\right).
\end{equation}

\noindent Comparing with \eqref{our entropy formula} we see that while our entropy formula depends on the eigenvalues of the connection in the direction of the non-contractible cycle of the torus, in the formalism of \cite{Gutperle:2011kf,Ammon:2011nk} it instead depends on the eigenvalues of the holomorphic and anti-holomorphic components of the connection, suggesting a form of holomorphic factorization. To further support this picture, using the smoothness conditions and repeating the arguments that led to \eqref{variation entropy}, in the present case we arrive at
\begin{align}\label{variation GK entropy}
\delta S_{\text{G-K}} 
={}&
    -2\pi i k_{cs}\text{Tr}\Biggl[ \tau\, \delta\left( \frac{a_{z}^{2}}{2}  \right)-\bar{\tau}\,\delta \left( \frac{\bar{a}_{\bar{z}}^{2}}{2}\right)
    +\bar{\tau}a_{\bar{z}} \,\delta \left(a_{z}-\Lambda^{+}\right) +\tau\bar{a}_{z}\,\delta\left(-\bar{a}_{\bar{z}}+\Lambda^{-}\right)\Biggr].
%    \nonumber\\
%    ={}&
%     -2\pi i k_{cs}\Bigl[ \tau\, \delta T-\bar{\tau}\,\delta \bar{T}+\sum_{j=3}^{N}\left(\mu_{j}\delta Q_{j}-\bar{\mu}_{j}\delta\bar{Q}_{j}\right)\Bigr]
\end{align}

\noindent We see that both the entropy \eqref{our holomorphic formula} and its variation \eqref{variation GK entropy} will have the correct form provided one identifies the stress tensor modes as purely  holomorphic/anti-holomorphic:
\begin{equation}
T_{\text{G-K}} = \text{Tr}\left[\frac{a_{z}^{2}}{2}\right]\,,\quad \bar{T}_{\text{G-K}} = \text{Tr}\left[\frac{\bar{a}_{\bar{z}}^{2}}{2}\right],
\end{equation}

\noindent as opposed to \eqref{stress tensor formula}, and demand that the lowest (highest) weights in $\bar{\tau}a_{\bar{z}}$ ($-\tau\bar{a}_{z}$) are linear in the higher spin chemical potentials, i.e.
\begin{align}\label{Euclidean GK sources and vevs}
\text{Tr}\left[\left(a_{z}-\Lambda^{+}\right)\bar{\tau}a_{\bar{z}}\right] ={}&
 \text{Tr}\bigl[Q_{\text{G-K}}\left(\bar{\tau}a_{\bar{z}}\right)\bigr] = \sum_{j=3}^{N}\left(\mu_{j}Q_{j}\right)_{\text{G-K}}
 \\
 \text{Tr}\left[\left(-\bar{a}_{\bar{z}}+\Lambda^{-}\right)\left(-\tau\bar{a}_{z}\right)\right]  ={}&
   \text{Tr}\bigl[\bar{Q}_{\text{G-K}}\left(-\tau\bar{a}_{z}\right)\bigr] = \sum_{j=3}^{N}\left(\bar{\mu}_{j}\bar{Q}_{j}\right)_{\text{G-K}}\,,
\label{Euclidean GK sources and vevs 2}
\end{align}

\noindent with the remaining ambiguity fixed by 
\begin{align}
\text{Tr}\Bigl[\Lambda^{+}\bar{\tau}a_{\bar{z}}\Bigr] ={}& \sum_{j =3}^{N}(j-1)\left(\mu_{j}Q_{j}\right)_{\text{G-K}}
\\
\text{Tr}\Bigl[\Lambda^{-}\tau\bar{a}_{z}\Bigr] ={}& \sum_{j =3}^{N}(j-1)\left(\bar{\mu}_{j}\bar{Q}_{j}\right)_{\text{G-K}}\,.
\end{align}

\noindent These expressions should be contrasted with those appropriate to the ``canonical" formalism, namely \eqref{Euclidean normalizations}-\eqref{Euclidean normalizations 1b} and \eqref{Euclidean normalizations 2a}-\eqref{Euclidean normalizations 2b}. Following the same logic as in \ref{section: Thermo}, one can construct appropriate boundary terms that enforce the holomorphic boundary conditions, and the entropy is then obtained as the Legendre transform of the free energy, which in this case reads
\begin{align}\label{holomorphic free energy}
-\beta F_{\text{G-K}} =
 \ln Z_{\text{G-K}}
  ={}&
   -2\pi i k_{cs}\text{Tr}\biggl[\tau\left(\frac{a_{z}^{2}}{2} \right) -\bar{\tau}\left( \frac{\bar{a}_{\bar{z}}^{2}}{2}\right)
     +\left(\bar{\tau} \Lambda^+ a_{\bar{z}}-\tau\Lambda^{-} \,\bar{a}_{z}\right)\biggr].
\end{align}

A potentially convenient way to summarize the relation between the holomorphic and canonical approaches is to notice that they are connected by a field redefinition: if $c$ denotes the connection appropriate to the holomorphic formalism, we write 
\begin{equation}
a_{z} = c_{z} -\frac{\bar{\tau}}{\bar{\tau}-\tau}c_{\bar{z}}\,,\qquad a_{\bar{z}} = \frac{\bar{\tau}}{\bar{\tau}-\tau}c_{\bar{z}}\,,
\end{equation}

\noindent and similarly for the barred connection, where the boundary conditions are now
\begin{equation}
c_{z} = \Lambda^{+} + Q\,,\qquad c_{\bar{z}} = M +\ldots 
\end{equation}

\noindent and the normalization is
\begin{equation}
\mbox{Tr}\left[\left(c_{z}-\Lambda^{+}\right)\bar{\tau}c_{\bar{z}}\right] = \sum_{j=3}^{N}\mu_{j}Q_{j}\,.
\end{equation}

%%%%%%%%%%%%%%%%%%%%%%%%%%%%%%%%%%%%%%%%%%%%%%%
\subsection{The connection to $2d$ CFT}
%%%%%%%%%%%%%%%%%%%%%%%%%%%%%%%%%%%%%%%%%%%%%%%

It is peculiar that we have found two different versions of Chern-Simons theory that differ only in their choice of boundary terms and conjugate variables, but otherwise both seem intimately connected to $2d$ conformal field theories with higher spin symmetry. In fact, we have at least three different versions, as one could also have decided to work with a torus with fixed periodicities and trade off $\tau$ and $\bar{\tau}$ for extra contributions in $a_{\bar{z}}$ and $\bar{a}_z$ as we did in section~\ref{HS gravity}. It is the latter one whose connection to CFT we understand quite well in view of \cite{deBoer:1991jc,DeBoer:1992vm}, where it was shown that the flatness conditions for connections in Drinfeld-Sokolov form are identical to the Ward identities of a CFT deformed by $\int d^2 z \sum_j \mu_j(z,\bar{z}) Q_j(z,\bar{z})$. Though these Ward identities were derived on the complex plane, the are approximately valid on a torus as well. 

The partition function for constant connections in Drinfeld-Sokolov form on a torus of \textit{fixed} periodicity $2\pi$ in each directions (i.e. $\text{Vol}(\partial M) = 4\pi^{2}$), in this formalism, is equal to
\begin{align} \label{jan1}
\ln Z_{\text{fixed}} = 2\pi k_{cs} {\rm Tr}&\Biggl[ \left(\frac{a_z^2}{2} - a_z a_{\bar{z}} - \frac{a_{\bar{z}}^2}{2}\right) 
+ 2 a_{\bar{z}} \left(a_{z}-\Lambda^+ \right) 
\nonumber\\
 &
- \left(\frac{\bar{a}_z^2}{2} +\bar{a}_z \bar{a}_{\bar{z}} - \frac{\bar{a}_{\bar{z}}^2}{2}\right) -2 \bar{a}_z\left(-\bar{a}_{\bar{z}}+\Lambda^- \right) \Biggr].
\end{align}
This is therefore the only version of the free energy whose a priori connection to CFT variables we understand quite well. Upon Legendre-transforming, the corresponding entropy is found to be
\begin{equation}
\label{jan2}
S_{\text{fixed}} = 2\pi k_{cs} {\rm Tr} \left[ \left(\frac{a_z^2}{2} - a_z a_{\bar{z}} - \frac{a_{\bar{z}}^2}{2}\right) -
\left(\frac{\bar{a}_z^2}{2} + \bar{a}_z \bar{a}_{\bar{z}} - \frac{\bar{a}_{\bar{z}}^2}{2}\right)
 \right] .
\end{equation}
Let us now try to match this formalism to the holomorphic formalism. To do so, we introduce new connections $b_z$ and $b_{\bar{z}}$ such that
\begin{equation} \label{jan0}
a_z=b_z\,,\qquad a_{\bar{z}} = a_z +\frac{i}{2}\left(\tau b_z + \bar{\tau} b_{\bar{z}}\right),
\end{equation}

\noindent and with the further condition that $b_{\bar{z}}$ does not contain a contribution from a spin two chemical potential anymore. The spin two chemical potential has thus been traded off for the parameter $\tau$. For simplicity, we will ignore the contributions of $\bar{a}_z$ and $\bar{a}_{\bar{z}}$ in what follows but these connections can be treated in the same way. We will also add an extra boundary term
\begin{equation} \label{jan5}
I_{\rm extra} = -2\pi k_{cs}\left(1+\frac{i\tau}{2}\right) {\rm Tr}\left[ b_z(2\Lambda^+ - b_z)\right]
\end{equation}
to the action for reasons that will become clear shortly. With the extra boundary term included, the variation of $\ln Z$ expressed in terms of the variables $b_z$ and $b_{\bar{z}}$ reads
\begin{equation}
\label{jan3}
\delta \left(\ln Z_{\text{fixed}}-I_{\rm extra}\right) = 2 \pi\,i k_{cs} {\rm Tr}\left[\frac{b_z^2}{2} \, \delta \tau +   \delta \left( \bar{\tau} b_{\bar{z}}\right)\left(b_z - \Lambda^+ \right) \right]
\end{equation}
which displays exactly the sets of conjugate variables of the holomorphic formalism. Moreover, due to \eqref{jan0}, the trivial holonomy condition for $a_z-a_{\bar{z}}$ is replaced by a condition on the holonomy of $\tau b_z + \bar{\tau} b_{\bar{z}}\,$, which once more agrees with the holomorphic formalism. If we subsequently compute the entropy by the appropriate Legendre transform we find that, in the presence of the extra boundary term,
\begin{equation}
\label{jan4}
\tilde{S}_{\text{fixed}}=-2\pi i k_{cs} {\rm Tr} \left[ b_z \left(\tau b_z + \bar{\tau} b_{\bar{z}}\right) 
+\frac{i}{8} \left(\tau b_z + \bar{\tau} b_{\bar{z}}\right)^2\right],
\end{equation}

\noindent plus a similar contribution from the other $SL(N,\mathds{R})$ gauge field. This is almost identical to the entropy given in \ref{our holomorphic formula}, except for the second term. This second term, however, is a constant which turns out to be equal to $\pi c /48\,$, and we can easily shift the definition of the free energy to get rid of this constant. We have therefore shown that if we start with the known Chern-Simons formulation of the CFT partition function in the presence of higher spin operators from 20 years ago and make a mild field redefinition we obtain precisely the entropy in the holomorphic formulation. In the present context, this observation partially explains why the CFT computations of \cite{Kraus:2011ds,Gaberdiel:2012yb} are in precise agreement with the entropy as computed in the holomorphic formulation, rather than its canonical version.

We also note that the necessity for the extra boundary term in \eqref{jan5} is related to the fact that in \cite{deBoer:1991jc,DeBoer:1992vm} the spin-two chemical potential couples to ${\rm Tr}\left[\Lambda^+ a_z\right]$ whereas in the holomorphic formalism $\tau$ couples to $\frac{1}{2} {\rm Tr} \left[a_z^2\right]\,$. These two operators are the same for the principal embedding and the extra boundary term vanishes in those cases, but for other embeddings the two operators differ in their contribution from the residual current algebra. The extra term in \eqref{jan5} precisely accomplishes the correct shift in the operator dual the spin-two chemical potential, or $\tau\,$.

What is left is to explain the connection between CFT computations and the canonical formulation studied in this paper. This will probably require us to find the appropriate change of variables and additional boundary terms which map the fixed periodicity formalism with its known CFT interpretation to the canonical formalism. We leave this interesting question to future work.

%%%%%%%%%%%%%%%%%%%%%%%%%%%%%%%%%%%%%%%%%%%%%%%
\subsection{Conclusions}
%%%%%%%%%%%%%%%%%%%%%%%%%%%%%%%%%%%%%%%%%%%%%%%
In the context of higher spin theories in AdS$_{3}$ displaying asymptotic $\mathcal{W}_{N}$ symmetries, we have provided definitions for the free energy and entropy of black hole solutions that follow naturally from the Chern-Simons variational principle, and moreover agree with those computed via canonical methods. Our expression for the entropy applies to solutions with radial dependence given by \eqref{general form connections} (which can always be achieved using the gauge freedom), where $a$ and $\bar{a}$ are constant connections written in the so-called Drinfeld-Sokolov form. In particular, for black hole solutions that relax the AdS$_{3}$ asymptotics via the inclusion of irrelevant higher spin operators in the dual theory, we have indicated how to define the higher-spin chemical potentials (sources) and charges (EVs) in a way that is consistent with the canonical structure of the theory, and includes rotating solutions. An advantage of our approach is that the expressions for thermodynamics quantities are valid for any embedding and are  moreover given entirely in terms of the connection components (c.f. \eqref{stress tensor formula}, \eqref{canonical free energy}, \eqref{our entropy formula}), as appropriate to the topological character of the bulk theory, and therefore do not rely on specific details of how the solutions are parameterized other than the aforementioned gauge choices. Furthermore, we have shown that the entropy of black hole solutions corresponding to constant connections can be written very compactly in terms of the eigenvalues of the connection in the direction of the non-contractible cycle of the Euclidean torus (c.f. \eqref{our entropy formula v2}).

An interesting feature of higher spin thermodynamics in this canonical formalism is that the black hole energy and entropy receive contributions that are non-linear in the chemical potentials and charges.  A related observation is that the natural boundary conditions in the higher spin theory introduce a coupling between left- and right-movers, breaking holomorphic factorization. We have shown that one can transition to a formalism that preserves the holomorphic structure via a field redefinition, and it would be of interest to explore this relation in more detail. This might also allow us to understand what the equivalent CFT
computations of the canonical formalism are. 

Although we used the term ``canonical formalism'' to describe the formalism with naive analytically continued boundary terms on the Euclidean torus, from a Chern-Simons perspective this formalism is not that canonical at all. Since Chern-Simons theory is a topological theory, the topology of the torus is all that matters in principle, and it is only the boundary terms that break the topological nature of the theory. The holomorphic formulation employs related but not identical boundary conditions and boundary terms. One can therefore not unambiguously discuss notions of energy without properly specifying boundary terms and boundary conditions in Chern-Simons theory. 

We have given very simple expressions for the entropy in terms of eigenvalues of the constant gauge field $a$ for both the canonical and the holomorphic formulation. Since the gauge field $a_z$ contains only charges, in the holomorphic formalism we do not have to solve any monodromy conditions in order to find the entropy: the only thing we need to do is to find the eigenvalues of $a_z$ and then take suitable linear combinations. Similar comments apply to the canonical formalism, because, for smooth constant connections, the $a_{\varphi}$ component is related in a simple way to $a_{z}\,$.  For the holomorphic formalism we have demonstrated that the entropy agrees with the corresponding CFT computation, and the corresponding linear combination of the eigenvalues can therefore be viewed as the generalization of the Cardy formula to higher spin algebras. What is not clear, and would be interesting to explore, is the range of validity of these answers. The partition function that Chern-Simons theory computes is that of a CFT on the plane deformed by higher spin operators. Once we replace the plane by a torus, the Ward identities will be modified in a way which in principle depends on the details of the CFT, and this will also affect the detailed form of the entropy. Similarly, quantum effects will modify the form of the entropy. Thus one expects the results to be valid in a high-temperature regime for not too large values of the higher spin chemical potentials, but it would clearly be desirable to have a more precise statement. 

There has been some discussion in the literature regarding the interpretation of these higher spin black holes. Since they correspond to a theory deformed by irrelevant operators, they no longer describe ``asymptotically anti-de Sitter" spacetimes in the strict sense. In our opinion, this is not a problem, since even in the presence of higher spin chemical potentials there are perfectly reasonable boundary conditions that one can impose. If one would ignore the issue of boundary terms and boundary conditions, a flat gauge field with trivial holonomy around the contractible cycle can be gauge-equivalent to gauge fields in various inequivalent higher spin theories. After all, the only invariant data in the flat gauge field is the holonomy around the non-contractible cycle, and this holonomy is precisely what parametrizes the higher spin black holes in the various higher spin theories. However, once boundary terms and boundary conditions are included, the resulting black holes are not physically equivalent to each other, even if the gauge fields that describe them are gauge-equivalent. They describe different thermal states in different conformal field theories with different asymptotic symmetry groups. If the constant gauge fields on the boundary for two different solutions are conjugate to each other their entropies will be equal, but if only the holonomy around the non-contractible cycle is identical entropies in different higher spin theories will correspond to different linear combinations of the eigenvalues of the holonomy matrix. 

To briefly summarize the results of this paper once more: we have discussed various choices of boundary conditions and boundary terms in Chern-Simons theory, which are relevant for the study of higher spin theories and their holographic duals. We have explained which one yields results which agrees with canonical methods, and also which one is equivalent to existing CFT computations. For either case have we described the precise form of the energy, and provided very simple expressions for the entropy and free energy that include the general rotating case. The main remaining gap in our understanding is the precise relation between the canonical formulation and CFT computations, and we hope to get back to this issue in the near future. Another direction that deserves further study is the discussion of thermodynamics in the presence of bulk fields which are charged under the (Abelian and non-Abelian) spin-one fields present in all non-principal embeddings. Finally, this work is the result of various confusions we faced when trying to compute entanglement entropies in higher spin theories and comparing the results to the thermal entropy in the appropriate limit, and we hope to present our results for the entanglement entropy in the near future as well  \cite{deBoer:2013}.

%%%%%%%%%%%%%%%%%%%%%%%%%%%%%%%%%%%%%%%%%%%%%%%%%%%%%%%%%%%%%%%%%%%%%%%%%%%%%%%%
%%%%%%%%%%%%%%%%%%%%%%%%%%%%%%%%%%%%%%%%%%%%%%%%%%%%%%%%%%%%%%%%%%%%%%%%%%%%%%%%
\vskip 1cm
\centerline{\bf Acknowledgments}
It is a pleasure to thank Martin Ammon, Max Ba\~nados, Alejandra Castro, Marc Henneaux, Nabil Iqbal, Tadashi Takayanagi and Tomonori Ugajin for helpful conversations and correspondence. This work is part of the research programme of the Foundation for Fundamental Research on Matter (FOM), which is part of the Netherlands Organization for Scientific Research (NWO).

%\bibliographystyle{uiuchept}
%\bibliography{HigherSpin}

\providecommand{\href}[2]{#2}\begingroup\raggedright\endgroup

\end{document}